\documentclass[11pt]{article}
\RequirePackage{amsthm,amsmath}
\RequirePackage{natbib}

\usepackage{amsgen,amsmath,amstext,amsbsy,amsopn,amssymb,latexsym,float}
\usepackage{authblk}
\usepackage{array}
\usepackage{cite}
\usepackage{multirow}
\usepackage{graphicx}
\usepackage{amssymb}
\usepackage{hyperref,xr}
\usepackage{mathtools}
\usepackage[dvipsnames,table,dvipsnames*, svgnames*, hyperref]{xcolor}
\usepackage{algorithm}
\usepackage{algpseudocode}
\usepackage{enumitem}
\usepackage{hyperref}

\newcommand{\ind}{\perp\!\!\!\!\perp}
\algrenewcommand\algorithmicrequire{\textbf{Input:}}
\algrenewcommand\algorithmicensure{\textbf{Output:}}
\makeatletter
\newcommand{\Statey}{\Statex\hspace*{\ALG@thistlm}}
\def\doo{\mbox{do}}

\newtheorem{theorem}{Theorem}
\newtheorem{corollary}{Corollary}
\newtheorem{proposition}{Proposition}

\graphicspath{{./figures/}}

\title{\textbf{Post-selection inference for causal effects after causal discovery}}
\author[1]{Ting-Hsuan Chang\thanks{Corresponding author: tc3255@cumc.columbia.edu}}
\author[2]{Zijian Guo\thanks{A large proportion of this research was conducted when Z. Guo was an associate professor at Rutgers University.}}
\author[1]{Daniel Malinsky}
\affil[1]{Department of Biostatistics, Columbia University}
\affil[2]{Center for Data Science, Zhejiang University}

\date{\today}

\begin{document}

\maketitle

\begin{abstract}
Algorithms for constraint-based causal discovery select graphical causal models among a space of possible candidates (e.g., all directed acyclic graphs) by executing a sequence of conditional independence tests. These may be used to inform the estimation of causal effects (e.g., average treatment effects) when there is uncertainty about which covariates ought to be adjusted for, or which variables act as confounders versus mediators. However, naively using the data twice, for model selection and estimation, would lead to invalid confidence intervals. Moreover, if the selected graph is incorrect, the inferential claims may apply to a selected functional that is distinct from the actual causal effect. We propose an approach to post-selection inference that is based on a resampling and screening procedure, which essentially performs causal discovery multiple times with randomly varying intermediate test statistics. Then, an estimate of the target causal effect and corresponding confidence sets are constructed from a union of individual graph-based estimates and intervals. We show that this construction has asymptotically correct coverage for the true causal effect parameter. Importantly, the guarantee holds for a fixed population-level effect, not a data-dependent or selection-dependent quantity. Most of our exposition focuses on the PC-algorithm for learning directed acyclic graphs and the multivariate Gaussian case for simplicity, but the approach is general and modular, so it may be used with other conditional independence based discovery algorithms and distributional families. 
\end{abstract}

\section{Introduction}\label{sec1}
Causal discovery algorithms exploit patterns of conditional independence in observed data to select graphical models whose nodes represent random variables, and directed edges represent direct causal relationships. The selected graph can then be used to inform the estimation of causal effects (e.g., average treatment effects) by identifying valid adjustment sets or distinguishing confounders from mediators of some relationship of interest \citep{maathuis2009estimating}. Recent work has also investigated how to exploit graphical structure to estimate a target causal effect more efficiently \citep{smucler2022efficient}. However, using the data twice, for model selection and subsequent causal effect estimation, without acknowledging the uncertainty in model selection may lead to invalid statistical inference for the estimated causal effect. Moreover, even using sample-splitting to separate the model selection and estimation steps runs a risk: if the selected graph is incorrect, the inferential claims may apply to a chosen functional that is distinct from the actual causal effect.

In this paper, we propose an approach to combining causal discovery with effect estimation to produce valid asymptotic inference for effects in settings where the underlying causal model is unknown. We focus primarily on the PC-algorithm, which estimates a Markov equivalence class of direct acyclic graphs (DAGs) under the assumption of causal sufficiency (i.e., no unmeasured confounders) \citep{spirtes2000causation}. The PC-algorithm selects a graphical causal model among a space of possible DAGs by executing a sequence of conditional independence tests and then propagating a series of orientation rules that determine edge directions. Many variations and improvements to the PC-algorithm have been described \citep{colombo2014order,harris2013pc,gretton2009nonlinear,chakraborty2022nonparametric,sondhi2019reduced}, but these all share the same fundamental ingredients: statistical tests of conditional independence determine the absence of edges, certain patterns of conditional independence imply (given background assumptions) some orientations via the ``collider rule,'' and perhaps additional orientations follow from these in conjunction with constraints on the space of allowable graphs (e.g., acyclicity). Once a graph is selected, causal effects of interest may be estimated by combining graphical identification results such as the well-known ``back-door criterion'' with preferred estimators \citep{pearl2009causality,maathuis2009estimating}. 

While there has been extensive literature on valid causal inference after model selection, most of the existing methods deal with the selection of nuisance models (propensity scores and outcome regressions) or subsets of adjustment variables from among a high-dimensional set that is a priori assumed to be sufficient for confounding control \citep{belloni2014inference, moosavi2023costs, cui2024selective}. Very few methods have been proposed to provide valid confidence intervals for the target causal effects after graph selection. \citet{strieder2021confidenceIC} proposed a test inversion approach to constructing confidence intervals in the context of bivariate linear causal models with homoscedastic errors. \citet{strieder2023confidence} generalized this approach to the multivariate case. In both works, the proposal relies on strong parametric assumptions and bespoke algorithms, whereas the approach we propose can be combined with a variety of existing constraint-based algorithms and independence tests. The work most closely related to ours is the method proposed by \citet{gradu2022valid}, which constructs valid confidence intervals after score-based graph selection. The basic idea of their ``noisy GES" (greedy equivalence search) approach is to bound the degree of dependence between the data and the learned graph by introducing random noise into the graph selection process. Here the graph selection proceeds by greedy optimization of a model fit score. Their target of inference, however, depends on the selected graph: they estimate parameters that are functionals of the selected graph rather than a selection-agnostic ``true'' causal effect. We describe this more formally below.

To fill a gap in the existing literature, we propose an approach to post-selection inference after constraint-based graph selection. As in \citet{belloni2014inference}, the target of inference is fixed: the average treatment effect of an exposure $X_i$ on an outcome $X_j$ in the ``true" causal structure. We assume that $X_i$ precedes $X_j$ in time, but allow the graphical structure to be otherwise unknown. Our method builds upon the resampling strategy introduced in \citet{xie2022repro} and \citet{guo2023robust}. Though most of our exposition focuses on the PC-algorithm for learning DAGs and the multivariate Gaussian case for simplicity, the approach is general and modular, so it can be used with other conditional independence-based discovery algorithms and (semi-)parametric distributional families.   

\section{Problem setup and preliminaries}\label{sec2}

\subsection{Causal graphs and target of inference}\label{sec2.1}
We first introduce the notations and formally set up the causal query. Throughout this paper, we use $G=(V,E)$ to denote the DAG representing the true causal structure, where $V=\{1,...,d\}$ is the set of nodes and $E \subset V \times V$ is the set of directed edges. Let $\text{Adj}_i(G) \subset V$ and $\text{Pa}_i(G) \subset V$ denote the set of adjacencies and parents (direct causes) of node $i$ in $G$, respectively. Typically, one identifies a causal structure up to its Markov equivalence class, which is represented by a completed partially directed acyclic graph (CPDAG). The CPDAG implied by $G$, which we denote by $C$, is a mixed graph (with directed and undirected edges) representing a set of graphs all Markov equivalent to $G$. $C$ has the same adjacencies as $G$ and a directed edge $i \rightarrow j$ is in $C$ if and only if $i \rightarrow j$ is common to all DAGs Markov equivalent to $G$. 

Our query of interest is the average treatment effect of a particular exposure $i$ on a particular outcome $j$ in the true causal structure, denoted by $\beta_{i,j}(G)$, and it is known that $j$ is temporally later than $i$. We consider having access to a finite data set $\mathcal{D} = \{ X^{(k)} \}_{k=1}^n \equiv \{ (X_1^{(k)},...,X_d^{(k)}) \}_{k=1}^n$ of $n$ i.i.d. $d-$dimensional vectors from some joint distribution. Using the $\doo-$notation of Pearl, our target of inference can be written as
\begin{equation*}
    \beta_{i,j}(G) = E[X_j \mid \doo(X_i=x_i)] - E[X_j \mid \doo(X_i=x_i')],
\end{equation*}
where
\begin{equation*}
    E[X_j \mid \doo(X_i=x_i)] = \int E[X_j \mid x_i, X_{S(G)}] dX_{S(G)}
\end{equation*}
and $S(G) \subset V$ is a valid adjustment set for the effect of $i$ on $j$ in $G$. $E[X_j \mid \doo(X_i=x_i)]$ may be equivalently written as $E[X_j(x_i)]$ in the notation of potential outcomes and the adjustment functional corresponds to the well-known g-formula. By the DAG Markov properties, $\text{Pa}_i(G)$ always qualifies as a valid adjustment set \citep{pearl2009causality}. If $X_i$ is binary, $\beta_{i,j}(G)$ quantifies the average treatment effect of $X_i$ on $X_j$. If $X_i$ is continuous, this is a causal contrast between two fixed levels of exposure (e.g., low vs.\ high). In the linear-Gaussian setting discussed below, the parameter of interest describes a single ``unit'' increase in exposure ($x_i=x_i'+1)$ and corresponds simply to the regression coefficient for $X_i$ in a linear regression of $X_j$ on $X_i$ and $\text{Pa}_i(G)$. Generally the framework we describe may accommodate non- or semi-parametric estimators of average causal effects. For simplicity of notation, we omit the subscripts and use $\beta(G)$ to denote our target of inference in the following sections. 

\subsection{PC-algorithm allowing for temporal ordering}\label{sec2.2}
The PC-algorithm begins with a complete undirected graph (i.e., where all vertices are pairwise connected) and then executes a sequence of conditional independence tests to remove edges. The version of the PC-algorithm used in our work is described in Algorithm~\ref{alg1}. First, we use the ``PC-stable'' modification of the adjacency search introduced in \citet{colombo2014order}. (This introduces a global variable $\tilde{\text{Adj}}_i$ at the beginning of each iteration of the testing loop, which is only updated after all tests of a given conditioning cardinality are executed, thereby addressing an undesirable order-dependence in the original algorithm.) To improve the performance of PC, one may also incorporate partial background knowledge regarding the temporal ordering of the variables in cases where such information is available. Algorithm~\ref{alg1} incorporates possible temporal (or ``tiered'') ordering information by two modifications from the original PC-algorithm: (1) the conditional independence between variables $i$ and $j$ given conditioning set $S$ is not tested if any variable in $S$ lies in the future of both $i$ and $j$; (2) edges cannot be directed from a variable in a later tier/time point to a variable in an earlier tier/time point \citep{spirtes2000causation,andrews2021practical,petersen2021data}. We define $\text{O}(V)$ as a vector that specifies the partial temporal order of each node in $V$. For example, for a 3-node graph ($d=3$), $\text{O}(V)=(1,1,2)$ indicates that variables 1 and 2 (both in tier 1) are measured before variable 3 (in tier 2), thus excluding edges $3 \rightarrow 1$ and $3 \rightarrow 2$ from $G$. We denote by $\bar{\text{O}}_{ij}$ the set of nodes that are in a later tier than both $i$ and $j$. In the previous example, $\bar{\text{O}}_{12} = \{3\}$. The trivial ordering where all variables are in the same tier (i.e., no temporal ordering) imposes no modification to the original (or ``stable'') PC-algorithm. When the target of inference is $\beta_{i,j}(G)$, the effect of $i$ on $j$, a minimal ordering may simply impose that $j$ is in a later tier than $i$ and potentially all other variables.

The PC-algorithm may be used with any appropriate test of conditional independence. In the case where $X$ is multivariate Gaussian, conditional independence corresponds to zero partial correlation of zero (see Proposition 5.2 in \citet{lauritzen1996graphical}). 

\begin{proposition}
If $X = (X_1,...,X_d)$ is multivariate Gaussian, then for $i \neq j \in \{1,...,d\}$ and $S \subseteq \{1,...,d\} \setminus \{i,j\}$,  $X_i \ind X_j \mid X_S$ if and only if $\rho_{ij \mid S} = 0$, where $\rho_{ij \mid S}$ denotes the partial correlation between $X_i$ and $X_j$ given $X_S$.
\end{proposition}

We apply Fisher's z-transformation to the partial correlation for its variance stabilizing property:
\begin{equation*}
    Z(\rho_{ij \mid S}, n) = (n-|S|-3)^{1/2} \log \bigg(\frac{1+\rho_{ij \mid S}}{1-\rho_{ij \mid S}}\bigg),
\end{equation*}
where $|S|$ is the cardinality of the conditioning set. Under the null hypothesis $H_0: \rho_{ij \mid S}=0$, the Fisher's z-transformation of the sample partial correlation $Z(\hat{\rho}_{ij \mid S}, n)$ is asymptotically standard normal. Thus, for the conditional independence test (Test, $\alpha$) in Algorithm~\ref{alg1}, at significance level $\alpha$, we reject the null hypothesis if $|Z(\hat{\rho}_{ij \mid S}, n)| > \Phi(1-\alpha/2)$, where $\Phi(\cdot)$ denotes the standard normal cumulative distribution function. For other parametric or semi-parametric distributional families, other tests may be used \citep{shah2020hardness, xiang2020flexible, petersen2021testing, cai2022distribution}. 

\begin{algorithm}[H]
\caption{PC(Test, $\alpha$, $\text{O}(V)$) algorithm allowing for temporal ordering}\label{alg1}
\begin{algorithmic}[1]
\Require Samples of the vector $X = (X_1,...,X_d)$ and $\text{O}(V)$
\Ensure CPDAG $\widehat{C}$ 
\State Form the complete graph $\tilde{C}$ on node set $V$ with undirected edges.
\State Let $s = 0$
\Repeat
    \ForAll{$i \in V$}
        \State Define $\tilde{\text{Adj}}_i{(\tilde{C})} = \text{Adj}_i(\tilde{C})$
    \EndFor
    \ForAll{pairs of adjacent vertices $(i,j)$ s.t. 
    \Statey (1) $|\tilde{\text{Adj}}_i(\tilde{C})\setminus\{j, \bar{\text{O}}_{ij}\}| \geq s$ and 
    \Statey (2) subsets $S \subseteq |\tilde{\text{Adj}}_i(\tilde{C})\setminus\{j, \bar{\text{O}}_{ij}\}|$ s.t. $|S| = s$}
        \If{$X_i \ind X_j | X_S$ according to (Test, $\alpha$)}
            \State Delete edge $i-j$ from $\tilde{C}$. Save $S$ in $\text{sepset}_{i,j} = \text{sepset}_{j,i}$.  
        \EndIf
    \EndFor
    \State Let $s = s+1$
\Until{for each pairs of adjacent vertices $(i,j)$, $|\tilde{\text{Adj}}_i(\tilde{C})\setminus\{j, \bar{\text{O}}_{ij}\}| < s$} 
\State (a) Determine the v-structures, while respecting $\text{O}(V)$.
\State (b) Apply the Meek orientation rules, while respecting $\text{O}(V)$.
\State \Return $\widehat{C}$
\end{algorithmic}
\end{algorithm}

The orientation steps (a) and (b) can be described in the following way:
\begin{enumerate}[label=(\alph*)]
    \item For all triples $(i,k,j)$ such that $i \in \text{Adj}_k(\tilde{C})$ and $j \in \text{Adj}_k(\tilde{C})$ but $i \notin \text{Adj}_j(\tilde{C})$, orient $i - k - j$ as $i \rightarrow k \leftarrow j$ (called a v-structure or collider) if and only if $k \notin \text{sepset}_{i,j}$.
    \item After determining all v-structures, exhaustively apply the Meek rules (R1)-(R4) to orient remaining undirected edges \citep{meek1995causal}. 
\end{enumerate}

The first of these steps, sometimes called the ``collider rule,'' is the primary source of orientations in the PC-algorithm.\footnote{Alternatively, one can apply a modification called the majority rule introduced in  \citet{colombo2014order} to such triples. First, determine all subsets $S \subseteq \text{Adj}_i(\tilde{C})$ or $S \subseteq \text{Adj}_j(\tilde{C})$ that make $i$ and $j$ conditionally independent (called separating sets). $(i,k,j)$ is labeled as ambiguous if $k$ is in exactly 50\% of the separating sets; otherwise, it is labeled as unambiguous. If $(i,k,j)$ is unambiguous, orient $i - k - j$ as $i \rightarrow k \leftarrow j$ iff $k$ is in less than 50\% of the separating sets. The majority rule resolves the order-dependence issue, i.e., dependence on the order in which variables appear in the dataset, in the determination of v-structures.} The second step extends the graph to include additional orientations implied by the conjunction of the existing colliders and the assumption that the underlying graph structure is acyclic. In both steps, respecting the temporal ordering O$(V)$ means that 
edges between vertices $i$ and $j$ are always oriented (forced) $i \rightarrow j$ when $j$ is later than $i$ in the ordering.

In finite samples, multiple conditional independence tests may lead to incompatible edge orientations due to statistical testing errors, even if at a population level the data-generating distribution is Markov and faithful to a DAG. A common approach to handling this is to allow bi-directed edges $i \leftrightarrow j$ when colliders are ``discovered'' at both $i$ and $j$ \citep{colombo2014order}. The output may also contain directed cycles if multiple orientation decisions (at least one of which must be erroneous) imply a cycle. Graphs containing bi-directed edges or cycles are considered invalid in the sense that the arrowheads in the graph are not consistent with any DAG. Thus, the output of Algorithm~\ref{alg1} may not be a valid CPDAG in finite samples.

\section{A resampling-based approach to post-selection inference after causal discovery}\label{sec3}
In this section, we propose a method for constructing a confidence interval for $\beta(G)$ that has asymptotically correct coverage, addressing the post-selection problem following causal discovery. Our procedure consists of two steps. The first step essentially performs causal discovery multiple times with randomly varying intermediate test statistics. In the second step, the confidence interval is constructed from a union of individual intervals based on valid graphs generated in the first step. The theoretical justifications are given in Section~\ref{sec4}. We also briefly discuss how our method contrasts with the noisy GES method in \citet{gradu2022valid}.

\subsection{Step 1: Resampling and screening}
Algorithm~\ref{alg2} describes our resampling procedure. For the multivariate Gaussian case, within each of the $M$ repeated runs of the PC-algorithm (Algorithm~\ref{alg1}), we resample the test statistic for testing $H_0: \rho_{ij \mid S} = 0$ as follows, using a single draw from a Gaussian distribution centered on the sample partial correlation:
\begin{equation*}
    Z(\hat{\rho}^{[m]}_{ij \mid S},n) \overset{\text{i.i.d.}}{\sim} N(Z(\hat{\rho}_{ij \mid S},n), 1), \quad 1 \leq m \leq M.
\end{equation*}
We reject the null hypothesis if 
\begin{equation}
    |Z(\hat{\rho}_{ij \mid S}^{[m]},n)| > \tau(M) \cdot z_{\nu/2L},
\label{eq: lower threshold test}
\end{equation}
where $\nu \in (0, 1/2)$, $L = \frac{d(d-1)}{2} \times \big(\text{max}_{i \in V}|\text{Adj}_i(G)| + 1 \big)$, and $\tau(M) \in (0,1)$ is a shrinkage parameter that is a function of the number of resamples $M$. $\tau(M)$ reduces the independence decision threshold at an appropriate rate to ensure that the type I and type II errors shrink as $M \to \infty$; it thus plays a role analogous to the shrinking significance level $\alpha_n$ in the standard PC-algorithm, where $\alpha_n \to 0$ as $n \to \infty$ \citep{kalisch2007estimating}. $L$ is the upper bound on the total number of independence tests performed in a single run of the PC-algorithm. In practice, $L$ is often chosen by the user as a limit on the maximum size of conditioning sets considered.

\begin{algorithm}[H]
\caption{Executing the PC-algorithm multiple times using resampled test statistics}\label{alg2}
\begin{algorithmic}[1]
\Require Samples of the vector $X = (X_1,...,X_d)'$ and $\text{O}(V)$
\Ensure $M$ graphs 
\For {$m = 1,...,M$}
    \State Do Algorithm~\ref{alg1}, but using resampled test statistics for each test, see eq.~(\ref{eq: lower threshold test})
    \State \Return $\widehat{C}^{[m]}$
\EndFor
\State \Return $(\widehat{C}^{[1]},...,\widehat{C}^{[M]})$
\end{algorithmic}
\end{algorithm}

Among the $M$ resulting graphs from Algorithm~\ref{alg2}, if the $m$-th graph $\widehat{C}^{[m]}$ fails to be a valid CPDAG (e.g., containing cycles or bi-directed edges), it is viewed as distant from the CPDAG implied by $G$ and is thus discarded. Only the graphs that are valid CPDAGs are retained for computing a confidence interval, i.e., we keep the set
\begin{center}
    $\mathcal{M} = \{1 \leq m \leq M: \widehat{C}^{[m]}$ is a valid CPDAG\}.
\end{center}

\subsection{Step 2: Aggregation}
For each $m \in \mathcal{M}$, let $k_m$ be the number of DAGs represented by $\widehat{C}^{[m]}$. With $k_m$ DAGs ($\widehat{G}^{[m]}_1,...,\widehat{G}^{[m]}_{k_m}$), we use back-door adjustment in each DAG to obtain at most $k_m$ estimates of $\beta(G)$: 
\begin{center}
    $\hat{\beta}^{[m]}_k = \hat{\beta}(\widehat{G}^{[m]}_k), \quad 1 \leq k \leq k_m$,
\end{center}
as well as corresponding standard error estimates $\hat{\sigma}^{[m]}_{\beta_k}$, $1 \leq k \leq k_m$. Applying back-door adjustment to each DAG in a estimated equivalence class was proposed by \citet{maathuis2009estimating}, though they did not quantify statistical uncertainty in the resulting estimates using confidence intervals.

For a $(1-\gamma)\%$ confidence interval for $\beta(G)$, we first generate an interval for each $m \in \mathcal{M}$:
\begin{center}
    $\mbox{CI}^{[m]} = \mathop{\cup}\limits_{k=1,...,k_m} \bigg( \hat{\beta}^{[m]}_k - z_{\alpha_1/2}\hat{\sigma}^{[m]}_{\beta_k}, \hat{\beta}^{[m]}_k + z_{\alpha_1/2}\hat{\sigma}^{[m]}_{\beta_k} \bigg)$ 
\end{center}
where $\alpha_1 = \gamma - \nu$.
Then, the $(1-\gamma)\%$ confidence interval for $\beta(G)$ is taken as the union of the individual intervals:
\begin{center}
    $\text{CI}^{\text{re}} = \cup_{m \in \mathcal{M}} \mbox{CI}^{[m]}$.
\end{center}

For simplicity, our proposal implements the back-door adjustment procedure of \citet{maathuis2009estimating} -- which essentially focuses on possible graphical parents of the exposure as sufficient adjustment variables -- though this may not be the most efficient graph-based adjustment procedure. More efficient choices of adjustment set have also been described \citep{witte2020efficient,rotnitzky2020efficient,henckel2022graphical}, which may be straightforwardly combined with this proposal to narrow each individual confidence interval in the aggregated set.

We now highlight the key difference between our work and that of \citet{gradu2022valid} in terms of the causal query of interest. The target of inference in \citet{gradu2022valid} is the causal effect as a functional of the learned causal structure, whereas our approach directly targets the causal effect in the true unknown causal structure. Using the notations in this paper, the target estimand in \citet{gradu2022valid} is $\beta_{i,j}(\widehat{G})$ for variable pairs $(i,j) \subseteq [d] \times [d]$, where $\widehat{G}$ is the estimated graph (either a DAG or a CPDAG) from causal discovery. Their noisy GES method constructs a valid confidence interval for $\beta_{i,j}(\widehat{G})$, when the estimate $\hat{\beta}_{ij}(\widehat{G})$ is computed from the same data used to generate $\widehat{G}$. Depending on whether their selected $\widehat{G}$ agrees on adjustment sets for $(i,j)$ with the true $G$, their target parameter may or may not correspond to the true causal effect. In the case where there is disagreement, their target parameter may be viewed as a ``projection'' onto a working model $\widehat{G}$. In contrast, our resampling approach allows the construction of a confidence interval for the true causal effect $\beta_{i,j}(G)$ of any predetermined exposure-outcome pair $(i,j)$, independent of any model selection.

\section{Theoretical justification}\label{sec4}
In this section, we present the theorems that justify our approach in Section~\ref{sec3}. The proofs are provided in \hyperref[app1]{Appendix 1}.

As already mentioned, the PC-algorithm may be combined with various choices of conditional independence test statistic, with partial correlation being just one common choice in the Gaussian setting. In the theoretical exposition below, we use $\widehat{\psi}_{ij \mid S}$ to denote some arbitrary independence test statistic, which estimates some population-level quantity (measure of conditional association) $\psi_{ij \mid S}$ that vanishes under the null of conditional independence. For example, $\psi_{ij \mid S}$ may correspond to the conditional mutual information between $X_i$ and $X_j$ given $X_S$ or some other measure of distance between the joint and product densities: $f(x_i, x_j \mid x_S)$ versus $f(x_i \mid x_S)f(x_j \mid x_S)$. Others have proposed metrics based on the expected conditional covariance or using ideas from kernel regression or copula models \citep{shah2020hardness,xiang2020flexible,petersen2021testing,cai2022distribution}. What these and other proposal share is that they each define some estimator $\widehat{\psi}_{ij \mid S}$ for a functional $\psi_{ij \mid S}$ of the joint distribution that vanishes under the null of $X_i \ind X_j \mid X_S$. Our resampling procedure relies on the asymptotic normality of each individual estimator $\widehat{\psi}_{ij \mid S}$.

\begin{theorem}
\label{theorem1}  
For each $(i, j) \in [d] \times [d]$ and $S \subseteq \{1,...,d\} \setminus \{i,j\}$, let $\psi_{ij \mid S}$ be some functional of the true distribution $(X_1,...,X_d) \sim P$ such that $\psi_{ij \mid S} = 0$ if $X_i \ind X_j \mid X_S$. Suppose $\widehat{\psi}_{ij \mid S}$ satisfies
\begin{equation*}
    \mathop{\limsup}\limits_{n \rightarrow \infty} \textup{\textbf{P}}(|\widehat{\psi}_{ij \mid S} - \psi_{ij \mid S}| / \sigma(\widehat{\psi}_{ij \mid S}) \geq z_{\alpha/2}) \leq \alpha \quad \text{for} \quad  0<\alpha<1.
\end{equation*}
Then
\begin{equation*}
    \mathop{\liminf}\limits_{n \rightarrow \infty} \mathop{\liminf}\limits_{M \rightarrow \infty} \textup{\textbf{P}}\Big( \mathop{\min}\limits_{1 \leq m \leq M} \big\{ \mathop{\max}\limits_{i,j,S} \{ |\widehat{\psi}_{ij \mid S}^{[m]} - \psi_{ij \mid S}| / \sigma(\widehat{\psi}_{ij \mid S}) \} \big\} \leq err_n (M,\nu) \Big) \geq 1-\nu
\end{equation*}    
for any $0 \leq \nu \leq 1/2$ and $err_n(M,\nu)$ defined in the following \eqref{eq: err M defnition} such that $err_n(M,\nu) \rightarrow 0$ as $M \rightarrow \infty$. Each $\widehat{\psi}_{ij \mid S}^{[m]}$ is a single draw from a Gaussian distribution centered on $\widehat{\psi}_{ij \mid S}$:
\begin{equation*}
    \widehat{\psi}_{ij \mid S}^{[m]} \overset{\text{i.i.d.}}{\sim} N(\widehat{\psi}_{ij \mid S}, \sigma(\widehat{\psi}_{ij \mid S})), \quad 1 \leq m \leq M.
\end{equation*}
\end{theorem}

\begin{corollary}
\label{corollary1}
Suppose for each $(i, j) \in [d] \times [d]$
\begin{equation*}
     \mathop{\limsup}\limits_{n} \textup{\textbf{P}}(|Z(\hat{\rho}_{ij \mid S}) - Z(\rho_{ij \mid S})| \geq z_{\alpha/2}) \leq \alpha \quad \text{for} \quad 0<\alpha<1.
\end{equation*}
Then
\begin{equation*}
    \mathop{\liminf}\limits_{n} \mathop{\liminf}\limits_{M} \textup{\textbf{P}}\Big( \mathop{\min}\limits_{1 \leq m \leq M} \big\{ \mathop{\max}\limits_{i,j,S} \{ |Z(\hat{\rho}_{ij \mid S}^{[m]}) - Z(\rho_{ij \mid S})| \} \big\} \leq err_n (M,\nu) \Big) \geq 1-\nu
\end{equation*}    
for any $0 \leq \nu \leq 1/2$ and $err_n(M,\nu)$ defined as in Theorem~\ref{theorem1}. 
\end{corollary}

Corollary~\ref{corollary1} indicates that, with a sufficiently large resampling size $M$, there exists $1 \leq m^* \leq M$ such that the resampled partial correlations are almost the same as the true partial correlations with high probability. Corollary~\ref{corollary1} follows directly from Theorem~\ref{theorem1} in the special case of multivariate Gaussian data. In fact, Theorem~\ref{theorem1} can be adapted for other settings so long as the corresponding test statistic is asymptotically normal. (Test statistics may in general be correlated across tests, but as is evidenced from the proof, correlation across tests plays no role: the key property is just that some resampled test statistic is close to the true parameter.) Theorem~\ref{theorem2} uses the result of Theorem~\ref{theorem1} to establish the coverage property of $\text{CI}^{\text{re}}$. 

\begin{theorem}
\label{theorem2}
Suppose that the assumptions of Theorem~\ref{theorem1} and the following assumptions hold,
\begin{enumerate}
    \item the distribution $P$ of $X$ is Markov and faithful to DAG $G$, i.e., for $i \neq j \in \{1,...,d\}$ and $S \subseteq \{1,...,d\} \setminus \{i,j\}$,
        \begin{center}
            $X_i \ind X_j \mid X_S \iff$ node $i$ and node $j$ are d-separated by $S$ in $G$
        \end{center}
    \item the shrinkage parameter $\tau(M)$ satisfies
        \begin{center}
            $\tau(M) \cdot z_{\nu/2L} \geq err_n(M,\nu)$ and $\mathop{\lim}\limits_{M \rightarrow \infty}\tau(M) \cdot z_{\nu/2L}  = 0$
        \end{center}
    \item for any DAG $G'$,
        \begin{center}
            $\frac{1}{\sigma_\beta} \big(\hat{\beta}(G') - \beta(G')\big) \rightarrow_d N(0,1)$ and $\frac{\hat{\sigma}_\beta}{\sigma_\beta} \rightarrow_p 1$,
        \end{center}
    where $\sigma_\beta$ is the standard error of $\hat{\beta}(G')$ and $\hat{\sigma}_\beta$ is an estimate of $\sigma_\beta$.
\end{enumerate}
Then, the proposed CI satisfies
    \begin{equation*}
        \mathop{\liminf}\limits_{n \rightarrow \infty}\mathop{\liminf}\limits_{M \rightarrow \infty} \textup{\textbf{P}}(\beta \in \textup{CI}^{\textup{re}}) \geq 1-\gamma.
    \end{equation*}
\end{theorem}
The faithfulness requirement in Assumption 1 is a fundamental assumption in causal discovery, ensuring a one-to-one correspondence between conditional independence and the absence of edges \citep{spirtes2000causation}. If faithfulness is violated, conditional independence may incorrectly lead to the removal of an edge in the PC-algorithm. 

Following the proof of Theorem~\ref{theorem1} (see Appendix \hyperref[app1]{1}), we define the error term
\begin{equation}
    err_n(M, \nu) = \frac{1}{2} \bigg(\frac{2 \log n}{c(\nu)M}\bigg)^{1/L}, \quad \text{with} \quad c(\nu) = \bigg( \frac{1}{\sqrt{2\pi}} \bigg)^L \exp \bigg( -\frac{L}{2} (z_{\nu / 2L})^2 \bigg),
\label{eq: err M defnition}
\end{equation}
such that $err_n(M,\nu)$ tends to 0 with a sufficiently large $M$.
The above equation suggests that we choose the shrinkage parameter $\tau(M)$ in \eqref{eq: lower threshold test} as 
\begin{equation*}
    \tau(M) = c^* (\log n/M)^{1/L},
\end{equation*}
where $c^*$ is a positive constant. The logic behind this is that the shrinking threshold level in \eqref{eq: lower threshold test} is sufficient for us to recover the true DAG $G$ (or rather, the corresponding CPDAG $C$) for the $m^*$-th resample, where, as established in Theorem \ref{theorem1}, the $m^*$-th resampled partial correlations are almost the same as the true partial correlations. A larger $c^*$ tends to lead to a sparser graph by increasing the threshold for rejecting the null of conditional independence. A sparser graph may be missing more edges that correspond to confounding, which can lead to bias in the causal estimates and incorrect coverage. Increasing the sparsity may also decrease the chance of forming cycles or bi-directed edges that would make the graph invalid, potentially leading to more kept graphs (i.e., larger $|\mathcal{M}|$) in the screening step of our procedure. Hence, one can use the percentage of kept graphs, $(|\mathcal{M}|/M) \times 100\%$, to guide the choice of $c^*$. Our simulation results show that choosing the $c^*$ that minimizes the percentage of kept graphs can be a good choice (see more discussion in Section~\ref{sec5.2}). Therefore, in practice we suggest implementing the procedure iteratively, beginning with small $c^*$ values (e.g., 0.01) and increasing to find the $c^*$ that corresponds to the smallest $|\mathcal{M}|/M$. We assess the performance of $\text{CI}^{\text{re}}$ with different choices of $c^*$ in Section~\ref{sec5.2}. The choice of $\text{max}_{i \in V}|\text{Adj}_i(G)|$ (which affects the value of $L$) depends on the expected sparsity of the graph as well as computational constraints. The values of $L$ and $\nu$ are not crucial here, as $c^*$ can be expressed as a function of $L$ and $\nu$. Thus, we can vary the threshold for the test statistic by simply varying $c^*$.

Assumption 3 is a weak assumption on the chosen estimator of our target causal effect parameter, namely that the estimator is consistent and asymptotically normal for the true value of the parameter. Though the estimator is informed by a graph (i.e., using a graph-derived choice of adjustment variables), nothing relies on $G'$ being the true graph or any specific graph: the estimator is only required to converge to the corresponding adjustment functional $\int E[X_j \mid x_i, X_{S(G')}] dX_{S(G')} - \int E[X_j \mid x_i', X_{S(G')}] dX_{S(G')}$. This property is satisfied by most commonly used estimators of average treatment effects, i.e., estimators using the parametric g-formula, inverse probability weighting, doubly-robust estimators, etc.

Theorem~\ref{theorem2} shows that $\text{CI}^{\text{re}}$ is asymptotically valid under weak conditions. Another question that naturally arises is how conservative the interval can be expected to be, i.e., how much the proposed procedure ``inflates" interval length as compared to the unadjusted interval constructed assuming knowledge of the true DAG (the ``oracle'' interval). While this is difficult to characterize theoretically in general because it will depend in complex ways on various parameters of the true data-generating process that are unknown in practice, we can describe sufficient conditions such that $\text{CI}^{\text{re}}$ is asymptotically equivalent to the oracle interval. That is, we can state that so long as the non-zero associations are ``strong’’ enough such that the PC-algorithm can ``easily’’ estimate the correct graph, modifying PC with the resampling procedure does not come with any cost: the resulting interval will be with high probability the same as the oracle interval. 

\begin{theorem}
\label{theorem3}
Suppose that conditions of Theorem~\ref{theorem2} hold. For all $\psi_{ij \mid S} \neq 0$, if
\begin{equation}\label{eq:psi low bound} 
    |\psi_{ij \mid S}| > \Big( 2\sqrt{2\log n + 2\log M} + 2\log n \Big) \cdot \mathop{\max}\limits_{i,j,S}\big(\sigma(\widehat{\psi}_{ij \mid S}) \big),
\end{equation}
there is a sequence $\nu_n \to 0$ $(n \to \infty)$ such that the proposed CI satisfies
\begin{equation*}
    \mathop{\limsup}\limits_{n \rightarrow \infty}\textup{\textbf{P}}(\textup{CI}^{\textup{re}} = \textup{CI}^{\textup{oracle}}) = 1.
\end{equation*}
\end{theorem}

The key sufficient condition here is (\ref{eq:psi low bound}). This makes the true graph ``easy’’ to learn and so the resampling procedure just produces the same graph in every iteration. Our assumption thus amounts to an even stronger version of ``strong faithfulness’’ as studied in \citet{kalisch2007estimating} and \citet{uhler2013geometry}, and this assumption is indeed very strong, so the result is primarily of theoretical interest. The result also depends on a user-specified sequence for $\nu_n \to 0$ as $n \to \infty$ (whereas previously $\nu$ was fixed). In practice, it may be more informative to empirically examine the average length of the proposed interval in simulations, as we do in the next section.

\section{Simulations}\label{sec5}

\subsection{Data generation}\label{sec5.1}
We generate a 10-node random DAG $G$ where the expected number of neighbors per node, i.e., the expected sum of the in- and out-degree, is 7. We intentionally generate a dense graph, since dense graphs are generally harder to accurately recover using the PC-algorithm compared to sparse graphs. The edge weights are scaled to mitigate multicollinearity as follows. First, for each edge $i \rightarrow j$ in $G$, the weight $\tilde{w}_{ij}$ is drawn from a uniform distribution over $(-1, -0.5) \cup (0.5, 1)$. Then, the weights are scaled such that each variable would have unit variance if all its parents are i.i.d. standard normal \citep{mooij2020joint}:
\begin{center}
    $w_{ij} = \tilde{w}_{ij} / (\sum_{k \in pa_j(G)} \tilde{w}_{kj}^2 + 1)^{1/2}$.
\end{center}
We then simulate a $n$ i.i.d. copies of the 10-dimensional multivariate Gaussian data vector $X = (X_1,...,X_{10})$ according to $G$. For each variable $X_j \in \{X_1,...,X_{10}\}$,
\begin{center}
    $X_j = \sum_{k \in pa_j(G)} w_{kj} X_k + \epsilon_j$, \quad $\epsilon_j \sim N(0,1)$.
\end{center}

The simulated DAG is topologically ordered, so $j$ is a non-ancestor of $i$ for $i < j$. In the following, we assume partial knowledge of the temporal ordering of variables: $\text{O}(V)=(1,1,1,2,2,2,2,2,3,3)$. Our target estimand is the average treatment effect of variable 6 on variable 10, $\beta(G) = \beta_{6, 10}(G)$. 

\subsection{Choice of tuning parameter}\label{sec5.2}
Section~\ref{sec4} provided an expression for the shrinkage parameter $\tau(M) = c^* (\log n/M)^{1/L}$. To get a sense of appropriate $c^*$ choices, this section examines our method across different $c^*$ values. We implement the two-step procedure described in Section~\ref{sec3} to construct a $95\%$ confidence interval ($\text{CI}^{\text{re}}$) for $\beta(G)$. We fix the sample size $n = 500$ and let $\text{max}_{i \in V} |\text{Adj}_i(G)| = 7$ and $\nu = 0.025$. We consider realistic resampling sizes $M \in \{50, 100\}$, and for each value of $M$, consider $c^* \in \{0.006, 0.007, 0.008, 0.009, 0.01, 0.02, 0.03, 0.04\}$. The \texttt{R} package \texttt{tpc} is used to implement the PC-algorithm accounting for temporal ordering \citep{tpc}. We run 500 simulations for all combinations of $M$ and $c^*$. At the beginning of each simulation run, we generate a new random DAG $G$ and simulate data according to $G$ as outlined in Section~\ref{sec5.1}. 

For comparison purposes, in each simulation run, we also generate $95\%$ confidence intervals for $\beta(G)$ that do not account for the uncertainty in graph selection ($\text{CI}^{\text{naive}}$). Specifically, we implement Algorithm~\ref{alg1} with $\alpha \in \{0.01, 0.05\}$; if the output from Algorithm~\ref{alg1} is a valid graph $\widehat{C}$, we construct $\text{CI}^{\text{naive}}$:
\begin{center}
    $\text{CI}^{\text{naive}} = \mathop{\cup}\limits_{k} \hat{\beta}_k(\widehat{G}_k) \pm 1.96\hat{\sigma}^{\text{naive}}_{\beta_k}$,
\end{center}
where $\widehat{G}_k$ is a DAG in the Markov equivalence class represented by $\widehat{C}$; if the output is invalid, no confidence interval is computed. $\hat{\sigma}^{\text{naive}}_{\beta_k}$ is the standard error for the linear regression estimator that treats the selected graph as ``known.'' We also include 
\begin{center}
    $\text{CI}^{\text{oracle}} = \hat{\beta}(G) \pm 1.96\hat{\sigma}^{\text{oracle}}_\beta$ 
\end{center}
as a benchmark. $\hat{\beta}_{6,10}(G)$ can be obtained through a linear regression of $X_{10}$ on $X_{6}$, adjusting for $\text{Pa}_6(G)$, and by taking the coefficient corresponding to $X_6$. $\hat{\sigma}^{\text{oracle}}_\beta$ is the variance from this linear regression, now treating the true DAG $G$ as ``known." 

\begin{figure}
\centerline{\includegraphics[width=40 pc,height=32 pc]
{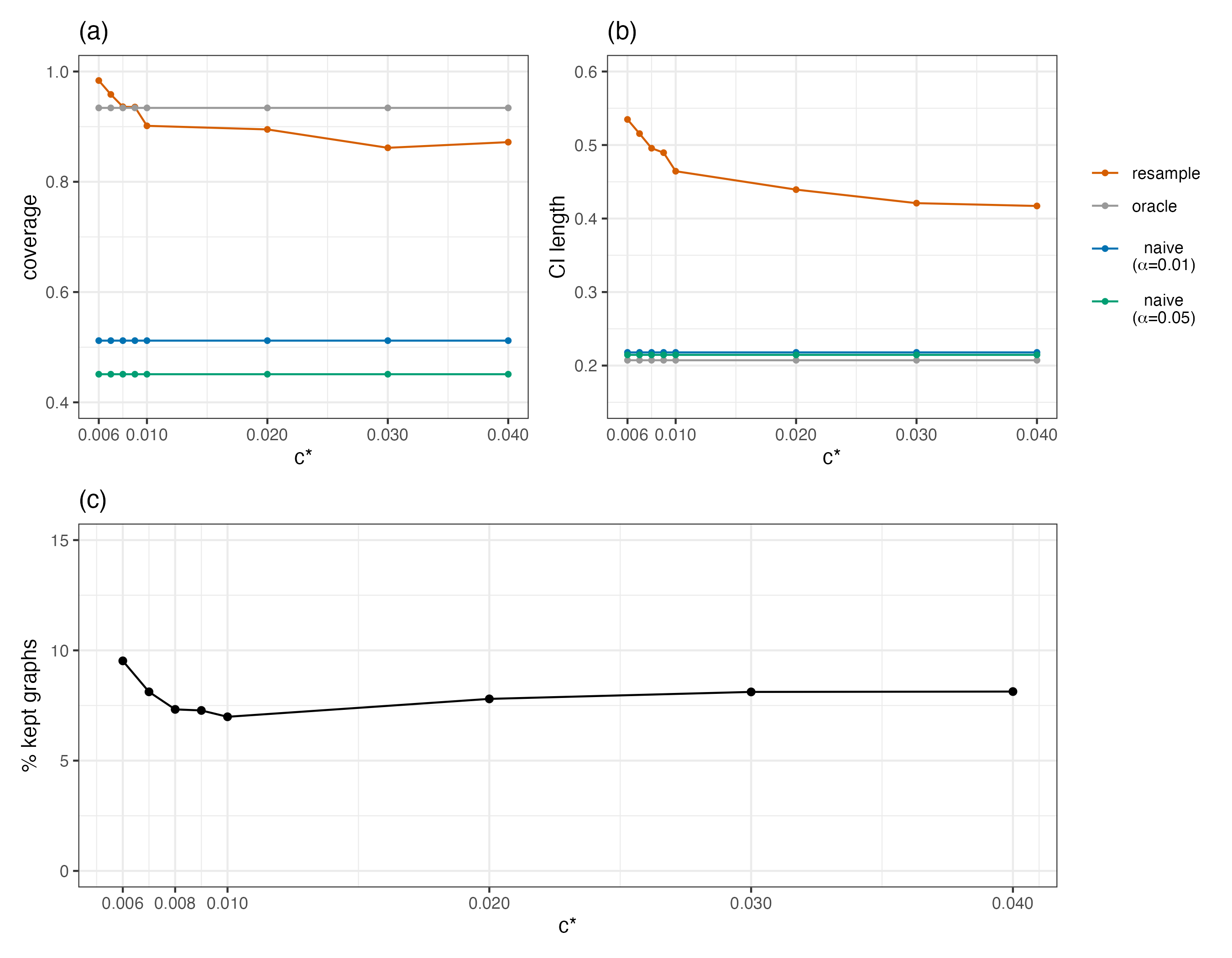}}
\caption{(a) Empirical coverage rate and (b) average length of $95\%$ CIs based on 500 simulations for varying $c^*$ values. ``naive" is the CI without incorporating the uncertainty in causal graph selection, ``oracle" is the CI with knowledge of the true causal graph, and ``resample" is our proposed CI in Section~\ref{sec3}, with $M=50$ resamples. (c) Percentage of valid graphs with $M=50$ resamples for constructing our proposed CI; results are based on 500 simulations for varying $c^*$. The true graph model possesses an average of 7 neighbors per node.}
\label{fig1}
\end{figure}

\begin{figure}
\centerline{\includegraphics[width=40 pc,height=32 pc]
{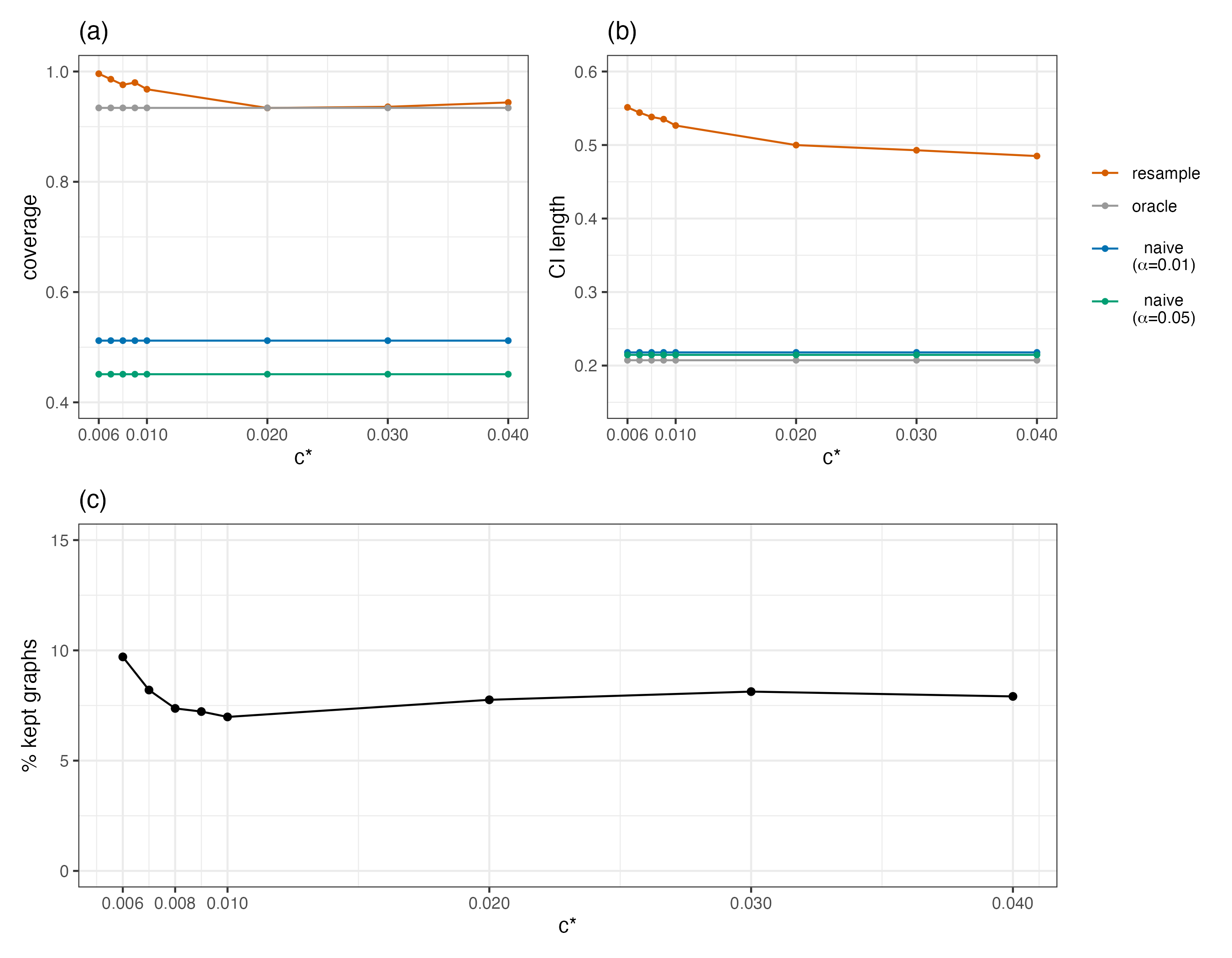}}
\caption{(a) Empirical coverage rate and (b) average length of $95\%$ CIs based on 500 simulations for varying $c^*$ values. ``naive" is the CI without incorporating the uncertainty in causal graph selection, ``oracle" is the CI with knowledge of the true causal graph, and ``resample" is our proposed CI in Section~\ref{sec3}, with $M=100$ resamples. (c) Percentage of valid graphs with $M=100$ resamples for constructing our proposed CI; results are based on 500 simulations for varying $c^*$. The true graph model possesses an average of 7 neighbors per node.}
\label{fig2}
\end{figure}

Figure~\ref{fig1} shows the empirical coverage rate and $95\%$ CI length from different methods, along with the percentage of valid graphs from the resampling method, based on 500 simulations for $M=50$ and varying $c^*$. Figure~\ref{fig2} shows these results for an increased resampling size of $M=100$. One can see that for a fixed $M$, the coverage rate of $\text{CI}^{\text{re}}$ tends to deteriorate as $c^*$ increases. As mentioned in Section~\ref{sec4}, a large $c^*$ may result in a sparse graph that misses the edges crucial for confounding adjustment in estimating $\beta(G)$. The increasing sparsity may also explain the increasing number of valid graphs observed from $c^* = 0.01$ to $c^* = 0.04$, since sparse graphs tend to have fewer cycles or bi-directed edges. Highly dense graphs at small $c^*$ values (e.g., 0.006) often have many undirected edges, hence less chance of having cycles or bi-directed edges, which may explain the slightly higher percentage of valid graphs from $c^* = 0.006$ to 0.009 versus $c^* = 0.01$. 

From both Fig.~\ref{fig1} and Fig.~\ref{fig2}, it appears that the coverage of $\text{CI}^{\text{re}}$ is generally close to $\text{CI}^{\text{oracle}}$ when the percentage of kept graphs is at its minimum (around $7\%$ in our case). Though choosing a very small $c^*$ (e.g., 0.006) that produces highly dense graphs typically ensures correct coverage, it is associated with reduced computational efficiency and potentially wider CIs. Thus, choosing a $c^*$ that minimizes the percentage of kept graphs represents a trade-off between computational efficiency and the validity of the effect estimate. By comparing Fig.~\ref{fig1} and Fig.~\ref{fig2}, we see that increasing $M$ improves the coverage of $\text{CI}^{\text{re}}$ and allows more tolerance for selecting larger $c^*$ values. This improvement, however, comes at the cost of computational speed and wider CIs. 

\subsection{Performance for different sample and resampling sizes}\label{sec5.3}
\begin{figure}
\centerline{\includegraphics[width=40 pc,height=32 pc]
{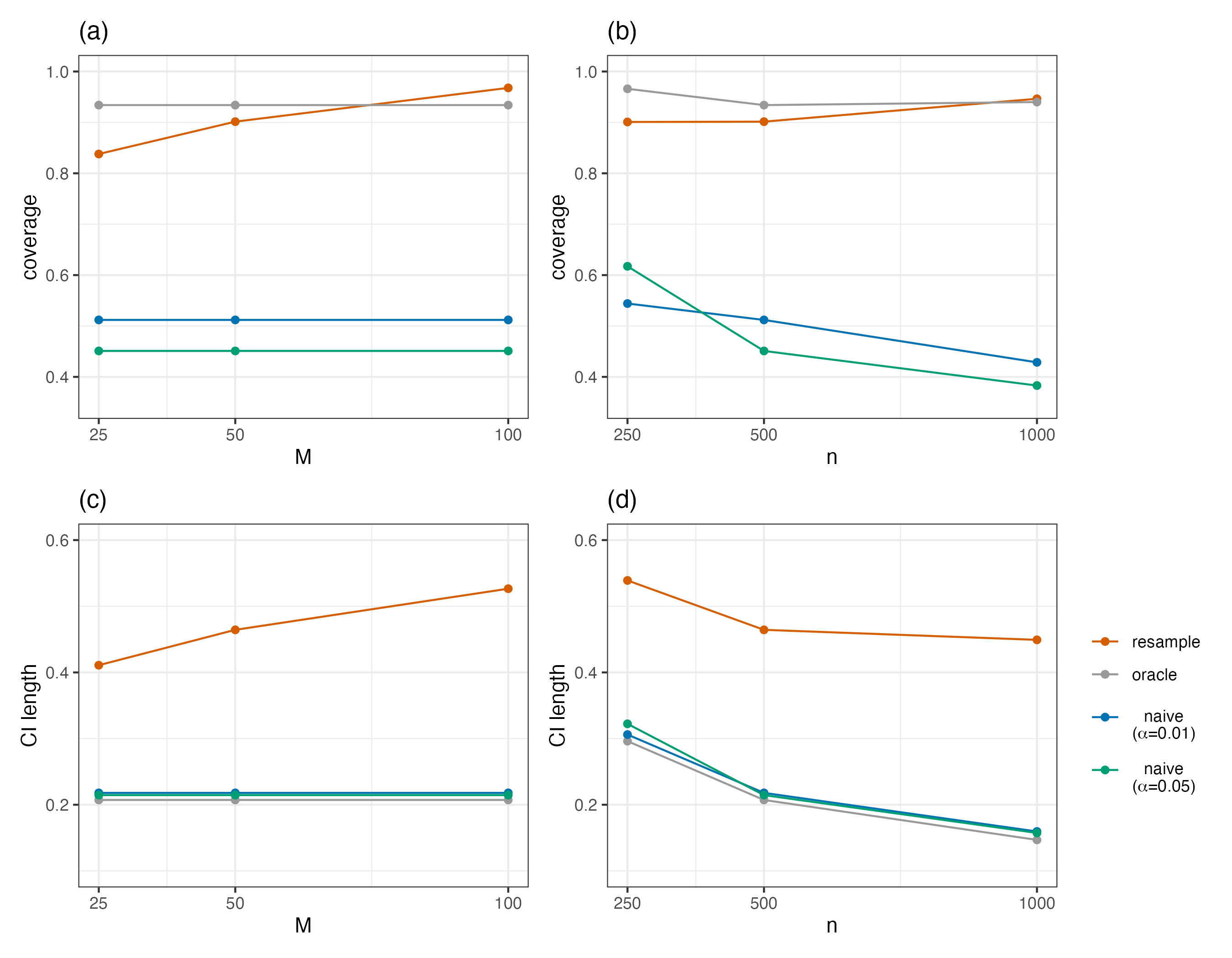}}
\caption{(a) Empirical coverage rate for varying resampling size $(M)$. (b) Empirical coverage rate for varying sampling size $(n)$. (c) Average $95\%$ CI length for varying $M$. (d) Average $95\%$ CI length for varying $n$. All results are based on 500 simulations. ``naive" is the CI without incorporating the uncertainty in causal graph selection, ``oracle" is the CI with knowledge of the true causal graph, and ``resample" is our proposed CI in Section~\ref{sec3}. The true graph model possesses an average of 7 neighbors per node.}
\label{fig3}
\end{figure}

Figure~\ref{fig3} compares the empirical coverage and average CI length between different types of CIs over 500 simulations for varying $n$ and $M$, fixing $c^* = 0.01$ in the resampling approach. As expected, the plots show an increase in both the coverage rate and length of $\text{CI}^{\text{re}}$ with increasing $M$. We observe that with an appropriate choice of $c^*$ and adequate number of resamples, $\text{CI}^{\text{re}}$ achieves correct coverage, whereas $\text{CI}^{\text{naive}}$ using either $\alpha = 0.01$ or $0.05$ exhibits less than $65\%$ coverage across all sample sizes considered. 

\subsection{Additional simulations}\label{sec5.4}
We additionally explore the performance of $\text{CI}^{\text{re}}$ with varying $c^*$ in the case of a sparse true DAG (the random DAG is generated such that the number of neighbors per node is 4); the results for $M=50$ are presented in Appendix \hyperref[app2.1]{2.1}. 

Another question we explore is whether the empirical performance of $\text{CI}^{\text{re}}$ may be driven by potential resamples from the tails of the Gaussian distribution (which might result in exhausting all possible graphs). To address this, we implement a truncated resampling approach, where each $Z(\hat{\rho}^{[m]}_{ij \mid S})$ is now drawn from a Gaussian distribution truncated at $\pm 1.5$ standard deviations from the sample test statistic $Z(\hat{\rho}_{ij \mid S})$. The results for $M=50$ are presented in Appendix \hyperref[app2.2]{2.2}, and they suggest that, at least empirically, the coverage and width of $\text{CI}^{\text{re}}$ are not sensitive to the resampling range.

Our proposed resampling approach resembles parametric bootstrapping at the level of individual independence test statistics. The nonparametric bootstrap, where the entire data set is resampled multiple times, may also be used to estimate a set of graph structures. However, naive combination of the nonparametric bootstrap with the PC-algorithm and estimation of causal effects does not necessarily lead to valid inference. We elaborate on this issue and evaluate the performance of a nonparametric bootstrap-based procedure in Appendix \hyperref[app2.3]{2.3}. 

Much of our exposition is framed in the linear‐Gaussian context. Nevertheless, the ideas are applicable in settings with nonlinear relationships and non-Gaussian distributions as well, and the PC‐algorithm can be combined with various choices of conditional independence test statistic, provided that the condition in Theorem~\ref{theorem1} is met. One nonparametric conditional independence test is based on the generalized covariance measure (GCM) statistic \citep{shah2020hardness}. We demonstrate our inference procedure combining the PC-algorithm with GCM in Appendix \hyperref[app2.4]{2.4}.

\section{Single-cell protein network data application}\label{sec6}
\citet{sachs2005causal} studied the performance of a structure learning procedure against a mostly ``known'' ground truth single-cell protein-signaling network. Their data included $n=7466$ concentration measurements of $d=11$ proteins: Plc, Pkc, Pka, PIP2, PIP3, Raf, Mek, Erk, Akt, P38, Jnk. We demonstrate our method using a clean version of the Sachs data provided in \citet{ramsey2018fask}. It has been suggested that the ground truth should include the directed edge Erk $\rightarrow$ Akt, since this causal relationship was confirmed by experimental manipulation \citep{sachs2005causal, ramsey2018fask}. Therefore, our goal is to construct a $95\%$ CI for the causal effect of Erk on Akt, under the assumption that Akt cannot be a parent of Erk. Based on the graphical model generated by \citet{sachs2005causal} and the supplemental ground truth described in \citet{ramsey2018fask}, we split the variables into two tiers: Plc, Pkc, Pka, PIP2 and PIP3 belong to tier 1, and the remaining variables belong to tier 2. Though the ground truth is thought to contain feedback loops (cycles), these are thought to occur mainly among variables in tier 1 and there are no directed edges expected from the variables in tier 2 toward variables in tier 1. Additionally, we assume that Raf and Mek are adjacent, as indicated in \citet{ramsey2018fask}, and that the relationships among the tier 2 variables can be represented by a DAG.

In Step 1 (resampling and screening), we let $\text{max}_{i \in V} |\text{Adj}_i(G)| = 7$, $\nu = 0.025$, and consider $c^* = 0.008, 0.01, 0.06$. We perform $M=100$ repeated runs of the tiered PC algorithm with tiers as specified earlier, and forbid the directed edge Akt $\rightarrow$ Erk to be included in the estimated graph. In the screening step, we keep a graph if it is a valid CPDAG among the tier 2 variables and if Raf and Mek are adjacent. Out of 100 estimated graphs, we keep 10, 6 and 9 graphs when $c^* = 0.008, 0.01$ and $0.06$, respectively. Thus, we choose $c^*=0.01$ since it corresponds to the minimum number of kept graphs. 

In Step 2 (aggregation), we orient the undirected component of tier 2 in each of the 6 kept graphs, leading to a total of 22 possibly different estimates of the target effect. We compare two versions for selecting the adjustment set from each model. The first version adjusts for all variables that are parents of Erk, resulting in an aggregated $95\%$ CI of (0.84, 0.94). Because all variables in tier 1 can be considered parents of Erk without introducing any bias, the second version adjusts for all tier 1 variables and parents of Erk in tier 2, resulting in an aggregated $95\%$ CI of (0.83, 0.93). The larger adjustment sets from the second version lead to slightly larger standard errors of the causal effect (mean = $7.80 \times 10^{-5}$), compared to those obtained from the first version (mean = $7.65 \times 10^{-5}$). In both cases, the estimated interval for the effect of interest is similar and does not include the null.

\section{Discussion and future work}\label{sec7}
In this work we introduce a resampling-based procedure for constructing asymptotically valid confidence intervals for causal effects after constrained-based causal discovery. While our focus has been on the PC-algorithm, our method has the potential to be adapted for other algorithms. For instance, the fast causal inference (FCI) algorithm \citep{spirtes2000causation} follows a very similar logic to the PC-algorithm to remove edges, but allows for unmeasured confounders with a more complicated set of orientation rules. Our proposed resampling procedure could be readily carried over to FCI. However, in settings with unmeasured confounders, the causal effect of interest may not be identified in some or all graphs in the produced equivalence class. Though there exist generalizations of the back-door adjustment strategy for estimating causal effects using the output of FCI, it is not so obvious how to aggregate a combination of numerical intervals with the agnostic result ``not identified.'' (One simple option would be to return the entire real line as the interval when some model suggests the effect is not identified; this would lead to trivially valid coverage but may not be desirable.) When a selected graph in the equivalence does not point-identify the causal effect of interest, it may imply bounds on the effect \citep{duarte2023automated, sachs2023general, gabriel2024sharp, bellot2024towards}. In that case, one possibility is to define lower and upper bounds as our inferential targets and construct valid confidence sets for these bounds instead of the causal effect. Extending our procedure to the context of unmeasured confounding is a potential direction for future work. We also believe that our resampling and screening procedure could be extended to permutation-based algorithms, e.g., the sparsest permutation algorithm \citep{raskutti2018learning} and the greedy sparsest permutation algorithm \citep{solus2021consistency} since these also use hypothesis tests of conditional independence, but the extension to entirely score-based algorithms is less clear. 

There are two important limitations in this work. Firstly, the user must choose a hyperparameter $c^*$ that controls the decision threshold for the conditional independence test. As we have shown, choosing a small value of $c^*$ will lead to correct coverage, but the width of the interval could be decreased if $c^*$ is chosen appropriately. We proposed one heuristic to choose this tuning parameter that appears to work well, but it would be beneficial if future work could find ways to improve on our heuristic or circumvent this issue. Secondly, the proposed approach is conservative and does not guarantee efficiency, i.e., the aggregated confidence interval can be wide. Though we found the interval width to be reasonable in our simulations and real data application, formal guarantees of ``minimal'' interval width inflation would be of interest.

\section*{Acknowledgements}

T. Chang and D. Malinsky were partially supported by the National Institutes of Health under award number R56AG082167 from the NIA, and D. Malinsky was also partially supported under award number K25ES034064 from NIEHS.
While at Rutgers University, the research of Z. Guo was partly supported by the NSF grant DMS 2015373 and NIH grants R01GM140463 and R01LM013614.

\newpage
\section*{Appendix 1: Proofs}\label{app1}

Our results build upon the theorems established in \citet{guo2023robust}.
\subsection*{Proof of Theorem~\ref{theorem1}}
\begin{proof}
Denote the observed data by $\mathcal{O}$, define
\begin{equation*}
    \mathcal{E} = \Big\{ \mathop{\max}\limits_{i,j,S} \{ |\widehat{\psi}_{ij \mid S} - \psi_{ij \mid S}|/\sigma(\widehat{\psi}_{ij \mid S}) \} \leq z_{\nu / 2L} \Big\}
\end{equation*}
where $L \equiv \frac{d(d-1)}{2} \times \big(\text{max}_{i \in V}|\text{Adj}_i(G)| + 1 \big)$.\\

\noindent Since $\mathop{\limsup}\limits_{n} \textbf{P}(|\widehat{\psi}_{ij \mid S} - \psi_{ij \mid S}|/\sigma(\widehat{\psi}_{ij \mid S}) \geq z_{\alpha/2}) \leq \alpha$, by applying the union bound, we have 
\begin{equation*}
    \mathop{\liminf}\limits_{n} \textbf{P}(\mathcal{E}) \geq 1-\nu.
\end{equation*}

\noindent Define $\hat{U}$ and $\{U^{[m]}\}_{1 \leq m \leq M} \in \mathbb{R}^{L}$ as
\begin{equation*}
    \hat{U} = \Bigg( \frac{\widehat{\psi}_{ij \mid S} - \psi_{ij \mid S}}{\sigma(\widehat{\psi}_{ij \mid S})} \Bigg)_{i,j,S} \text{ and } U^{[m]} = \Bigg( \frac{\widehat{\psi}_{ij \mid S} - \widehat{\psi}_{ij \mid S}^{[m]}}{\sigma(\widehat{\psi}_{ij \mid S})} \Bigg)_{i,j,S}.
\end{equation*}
Note that some of the partial correlations may never be estimated in a certain run of the PC-algorithm, but are nonetheless still well-defined.\\

\noindent The conditional density function of $U^{[m]}$ given the data $\mathcal{O}$ is
\begin{equation*}   
    f(U^{[m]} = U | \mathcal{O}) = \mathop{\Pi}\limits_{1 \leq l \leq L} \frac{1}{\sqrt{2\pi}} \exp(-\frac{U^2_l}{2})
\end{equation*}
since $\widehat{\psi}_{ij \mid S}^{[m]} \sim N(\widehat{\psi}_{ij \mid S}, \sigma(\widehat{\psi}_{ij \mid S}))$.\\

\noindent On the event $\mathcal{E}$, we have
\begin{equation*}
    \mathop{\sum}\limits_{l} \frac{\hat{U}^2_l}{2} \leq \frac{L}{2} (z_{\nu / 2L})^2
\end{equation*}
and further establish
\begin{equation}\label{eq:a1}
    f(U^{[m]} = \hat{U} | \mathcal{O}) \cdot 1_{\mathcal{O} \in \mathcal{E}} \geq c(\nu) := \big( \frac{1}{\sqrt{2\pi}} \big)^L \exp \big( -\frac{L}{2} (z_{\nu / 2L})^2 \big).
\end{equation}

\noindent Note that 
\begin{equation*}
\begin{split}
    & \textbf{P}\Big( \mathop{\min}\limits_{1 \leq m \leq M} ||U^{[m]} - \hat{U}||_\infty \leq err_n(M, \nu) | \mathcal{O} \Big) \\
    &= 1 - \textbf{P}\Big( \mathop{\min}\limits_{1 \leq m \leq M} ||U^{[m]} - \hat{U}||_\infty \geq err_n(M, \nu) | \mathcal{O} \Big) \\
    &= 1 - \mathop{\Pi}\limits^{M}_{m=1} \Big(1 - \textbf{P} \big(||U^{[m]} - \hat{U}||_\infty \leq err_n(M, \nu) | \mathcal{O} \big)\Big) \\
    &\geq 1 - \exp \Big( -\mathop{\sum}\limits_{m=1}^M \textbf{P} \big(||U^{[m]} - \hat{U}||_\infty \leq err_n(M, \nu) | \mathcal{O} \big) \Big),
\end{split}
\end{equation*}
where the 2nd equality follows from the conditional independence of $\{U^{[m]}\}_{1 \leq m \leq M}$ given $\mathcal{O}$ and the last inequality follows from $1-x \leq e^{-x}$.\\

\noindent By applying the above inequality, we establish
\begin{equation}\label{eq:a2}
\begin{split}
    & \textbf{P}\Big( \mathop{\min}\limits_{1 \leq m \leq M} ||U^{[m]} - \hat{U}||_\infty \leq err_n(M, \nu) | \mathcal{O} \Big) \cdot 1_{\mathcal{O} \in \mathcal{E}} \\ 
    &\geq \bigg( 1 - \exp \Big( -\mathop{\sum}\limits_{m=1}^M \textbf{P} \big(||U^{[m]} - \hat{U}||_\infty \leq err_n(M, \nu) | \mathcal{O} \big) \Big) \bigg) \cdot 1_{\mathcal{O} \in \mathcal{E}} \\
    &= 1 - \exp \Big( -\mathop{\sum}\limits_{m=1}^M \textbf{P} \big(||U^{[m]} - \hat{U}||_\infty \leq err_n(M, \nu) | \mathcal{O} \big) \cdot 1_{\mathcal{O} \in \mathcal{E}} \Big)
\end{split}
\end{equation}

\noindent Thus, for the remainder of the proof, we want to establish a lower bound for 
\begin{equation*}
    \textbf{P} \big(||U^{[m]} - \hat{U}||_\infty \leq err_n(M, \nu) | \mathcal{O} \big) \cdot 1_{\mathcal{O} \in \mathcal{E}},
\end{equation*}
and then establish the lower bound for 
\begin{equation*}
    \textbf{P}\Big( \mathop{\min}\limits_{1 \leq m \leq M} ||U^{[m]} - \hat{U}||_\infty \leq err_n(M, \nu) | \mathcal{O} \Big).
\end{equation*}

\noindent First, decompose
\begin{equation*}
\begin{split}
    & \textbf{P} \big(||U^{[m]} - \hat{U}||_\infty \leq err_n(M, \nu) | \mathcal{O} \big) \cdot 1_{\{\mathcal{O} \in \mathcal{E}\}} \\
    &= \int f(U^{[m]} = U | \mathcal{O}) \cdot 1_{\{||U - \hat{U}||_\infty \leq err_n(M, \nu)\}} dU \cdot 1_{\mathcal{O} \in \mathcal{E}} \\
    &= \int f(U^{[m]} = \hat{U} | \mathcal{O}) \cdot 1_{\{||U - \hat{U}||_\infty \leq err_n(M, \nu)\}} dU \cdot 1_{\mathcal{O} \in \mathcal{E}} \\
    &\hspace{3 mm} + \int \Big(f(U^{[m]} = U | \mathcal{O}) - f(U^{[m]} = \hat{U} | \mathcal{O})\Big) \cdot 1_{\{||U - \hat{U}||_\infty \leq err_n(M, \nu)\}} dU \cdot 1_{\mathcal{O} \in \mathcal{E}}
\end{split}
\end{equation*}

\noindent By (\ref{eq:a1}) we get
\begin{equation*}
\begin{split}
    & \int f(U^{[m]} = \hat{U} | \mathcal{O}) \cdot 1_{\{||U - \hat{U}||_\infty \leq err_n(M, \nu)\}} dU \cdot 1_{\mathcal{O} \in \mathcal{E}} \\
    &\geq c(\nu) \cdot \int 1_{\{||U - \hat{U}||_\infty \leq err_n(M, \nu)\}} dU \cdot 1_{\mathcal{O} \in \mathcal{E}} \\
    &\geq c(\nu) \cdot \big(2err_n(M, \nu)\big)^L \cdot 1_{\mathcal{O} \in \mathcal{E}}
\end{split}
\end{equation*}

\noindent By the mean value theorem, there exists $t \in (0,1)$ s.t. 
\begin{equation*}
    f(U^{[m]} = U | \mathcal{O}) - f(U^{[m]} = \hat{U} | \mathcal{O}) = \big[\nabla f(\hat{U} + t(U - \hat{U}))\big]^\intercal (U - \hat{U}),
\end{equation*}
with 
\begin{equation*}
    \nabla f(U^*) \equiv (\frac{1}{\sqrt{2\pi}})^L \exp(-\frac{U^{*^\intercal} U^*}{2}) \big[-U^*\big]^\intercal.
\end{equation*}

\noindent Since $||\nabla f||_2$ is upper bounded, $\exists C>0$ s.t.
\begin{equation*}
\begin{split}
    &\big|f(U^{[m]} = U | \mathcal{O}) - f(U^{[m]} = \hat{U} | \mathcal{O})\big| \\
    &\leq ||\nabla f(\hat{U} + t(U - \hat{U}))||_2 ||U - \hat{U}||_2 \hspace{31pt}(\text{by Cauchy-Schwarz inequality})\\
    &\leq ||\nabla f(\hat{U} + t(U - \hat{U}))||_2 \sqrt{L} ||U - \hat{U}||_\infty \hspace{12pt}(\text{by } \|a\|_2\leq \sqrt{L}\|a\|_{\infty} \quad  \forall a\in \mathbb{R}^L)\\
    &\leq C \sqrt{L} ||U - \hat{U}||_\infty
\end{split}
\end{equation*}

\noindent Then, we obtain 
\begin{equation*}
\begin{split}
    & \Big| \int \Big(f(U^{[m]} = U | \mathcal{O}) - f(U^{[m]} = \hat{U} | \mathcal{O})\Big) \cdot 1_{\{||U - \hat{U}||_\infty \leq err_n(M, \nu)\}} dU \cdot 1_{\mathcal{O} \in \mathcal{E}} \Big| \\
    &\leq C \sqrt{L} \cdot err_n(M, \nu) \cdot \int 1_{\{||U - \hat{U}||_\infty \leq err_n(M, \nu)\}} dU \cdot 1_{\mathcal{O} \in \mathcal{E}} \\
    &= C \sqrt{L} \cdot err_n(M, \nu) \cdot \big(2err_n(M, \nu)\big)^L \cdot 1_{\mathcal{O} \in \mathcal{E}}
\end{split}
\end{equation*}

\noindent Since $err_n(M, \nu) \rightarrow 0$ and $c(\nu)$ is a positive constant, there exists a positive integer $M_0$ s.t.
\begin{equation*}
    C \sqrt{L} \cdot err_n(M, \nu) \leq \frac{1}{2} c(\nu) \quad \text{for  } M > M_0.
\end{equation*}

\noindent By combining 
\begin{equation*}
    \int f(U^{[m]} = \hat{U} | \mathcal{O}) \cdot 1_{\{||U - \hat{U}||_\infty \leq err_n(M, \nu)\}} dU \cdot 1_{\mathcal{O} \in \mathcal{E}} \geq c(\nu) \cdot \big(2err_n(M, \nu)\big)^L \cdot 1_{\mathcal{O} \in \mathcal{E}}
\end{equation*}
and
\begin{equation*}
\begin{split}
    & \Big| \int \big(f(U^{[m]} = U | \mathcal{O}) - f(U^{[m]} = \hat{U} | \mathcal{O})\big) \cdot 1_{\{||U - \hat{U}||_\infty \leq err_n(M, \nu)\}} dU \cdot 1_{\mathcal{O} \in \mathcal{E}} \Big| \\ 
    &\leq C \sqrt{L} \cdot err_n(M, \nu) \cdot \big(2err_n(M, \nu)\big)^L \cdot 1_{\mathcal{O} \in \mathcal{E}} \\
    &\leq \frac{1}{2}c(\nu) \cdot \big(2err_n(M, \nu)\big)^L \cdot 1_{\mathcal{O} \in \mathcal{E}} \quad \text{for  } M \geq M_0
\end{split}
\end{equation*}
we obtain that for $M > M_0$,
\begin{equation*}
\begin{split}
    & \textbf{P} \big(||U^{[m]} - \hat{U}||_\infty \leq err_n(M, \nu) | \mathcal{O} \big) \cdot 1_{\{\mathcal{O} \in \mathcal{E}\}} \\
    &= \int f(U^{[m]} = \hat{U} | \mathcal{O}) \cdot 1_{\{||U - \hat{U}||_\infty \leq err_n(M, \nu)\}} dU \cdot 1_{\mathcal{O} \in \mathcal{E}} \\
    &\hspace{3 mm} + \int \Big(f(U^{[m]} = U | \mathcal{O}) - f(U^{[m]} = \hat{U} | \mathcal{O})\Big) \cdot 1_{\{||U - \hat{U}||_\infty \leq err_n(M, \nu)\}} dU \cdot 1_{\mathcal{O} \in \mathcal{E}} \\
    &\geq \frac{1}{2}c(\nu) \cdot \big(2err_n(M, \nu)\big)^L \cdot 1_{\mathcal{O} \in \mathcal{E}}
\end{split}
\end{equation*}

\noindent Together with (\ref{eq:a2}), we establish that for $M \geq M_0$,
\begin{equation*}
\begin{split}
    & \textbf{P}\Big( \mathop{\min}\limits_{1 \leq m \leq M} ||U^{[m]} - \hat{U}||_\infty \leq err_n(M, \nu) | \mathcal{O} \Big) \cdot 1_{\mathcal{O} \in \mathcal{E}} \\
    &\geq 1 - \exp \Big( -M \cdot \frac{1}{2} c(\nu) \cdot \big(2err_n(M, \nu)\big)^L \cdot 1_{\mathcal{O} \in \mathcal{E}} \Big) \\
    &= \Big\{ 1- \exp \Big( -M \cdot \frac{1}{2} c(\nu) \cdot \big(2err_n(M, \nu)\big)^L \Big) \Big\} \cdot 1_{\mathcal{O} \in \mathcal{E}}
\end{split}
\end{equation*}

\noindent With $E_{\mathcal{O}}$ denoting the expectation taken w.r.t. the observed data $\mathcal{O}$, we establish that for $M \geq M_0$,
\begin{equation*}
\begin{split}
    & \textbf{P}\Big( \mathop{\min}\limits_{1 \leq m \leq M} ||U^{[m]} - \hat{U}||_\infty \leq err_n(M, \nu) \Big) \\
    &= E_{\mathcal{O}} \Big( \textbf{P}\big( \mathop{\min}\limits_{1 \leq m \leq M} ||U^{[m]} - \hat{U}||_\infty \leq err_n(M, \nu) | \mathcal{O} \big) \Big) \\
    &\geq E_{\mathcal{O}} \Big( \textbf{P}\big( \mathop{\min}\limits_{1 \leq m \leq M} ||U^{[m]} - \hat{U}||_\infty \leq err_n(M, \nu) | \mathcal{O} \big) \cdot 1_{\mathcal{O} \in \mathcal{E}} \Big) \\
    &\geq E_{\mathcal{O}} \Big( \Big\{ 1- \exp \big( -M \cdot \frac{1}{2} c(\nu) \cdot \big(2err_n(M, \nu)\big)^L \big) \Big\} \cdot 1_{\mathcal{O} \in \mathcal{E}} \Big)
\end{split}
\end{equation*}

\noindent By the definition $err_n(M, \nu) = \frac{1}{2} \big(\frac{2 \log n}{c(\nu)M}\big)^{1/L}$, we establish that for $M \geq M_0$,
\begin{equation*}
    \textbf{P}\Big( \mathop{\min}\limits_{1 \leq m \leq M} ||U^{[m]} - \hat{U}||_\infty \leq err_n(M, \nu) \Big) \geq (1-n^{-1}) \cdot \textbf{P}(\mathcal{E}).
\end{equation*}

\noindent We further apply $\mathop{\liminf}\limits_{n} \textbf{P}(\mathcal{E}) \geq 1-\nu$ and establish
\begin{equation*}
    \mathop{\liminf}\limits_{n \rightarrow \infty} \mathop{\lim}\limits_{M \rightarrow \infty} \textbf{P}\Big( \mathop{\min}\limits_{1 \leq m \leq M} ||U^{[m]} - \hat{U}||_\infty \leq err_n(M, \nu) \Big) \geq \textbf{P}(\mathcal{E}) \geq 1-\nu.
\end{equation*}
\end{proof}

\subsection*{Proof of Theorem~\ref{theorem2}}
\begin{proof}
Define the event
\begin{equation*}
    \mathcal{E}_1 = \Big\{ \mathop{\min}\limits_{1 \leq m \leq M}\mathop{\max}\limits_{i,j,S} \big\{ \big|\widehat{\psi}_{ij \mid S}^{[m]} - \psi_{ij \mid S}\big|/\sigma(\widehat{\psi}_{ij \mid S}) \big\} \leq err_n(M, \nu) \Big\}.
\end{equation*}

\noindent On the event $\mathcal{E}_1$, we use $m^*$ to denote the index s.t.
\begin{equation*}
    \mathop{\max}\limits_{i,j,S} \big\{\big|\widehat{\psi}_{ij \mid S}^{[m^*]} - \psi_{ij \mid S}\big|/\sigma(\widehat{\psi}_{ij \mid S})\big\} \leq err_n(M, \nu).
\end{equation*}

\noindent Let $\mathcal{E}_{i,j|S}$ denote the event that ``an error occurred when testing $\psi_{ij \mid S} = 0$".
Thus,
\begin{equation*}
\begin{split}
    &\textbf{P}(\text{an error occurs in the PC run that produces } \widehat{C}^{[m^*]})\\ 
    &\leq \textbf{P}\Big( \mathop{\cup}\limits_{i,j,S \subseteq \{1,...,d\} \setminus \{i,j\}} \mathcal{E}_{i,j|S} \Big)
\end{split}
\end{equation*}

\noindent Note that
\begin{equation*}
    \mathcal{E}_{i,j|S} = \mathcal{E}^I_{i,j|S} \cup \mathcal{E}^{II}_{i,j|S},
\end{equation*}
where 
\begin{align*}
    & \text{type I error } \mathcal{E}^I_{i,j|S}: \big|\widehat{\psi}_{ij \mid S}^{[m^*]}\big|/\sigma(\widehat{\psi}_{ij \mid S}) > \tau(M) \cdot z_{\nu/2L} \text{ and } \psi_{ij \mid S}=0, \\
    & \text{type II error } \mathcal{E}^{II}_{i,j|S}: \big|\widehat{\psi}_{ij \mid S}^{[m^*]}\big|/\sigma(\widehat{\psi}_{ij \mid S}) \leq \tau(M) \cdot z_{\nu/2L} \text{ and } \psi_{ij \mid S} \neq 0.
\end{align*}

\noindent Since $\mathop{\max}\limits_{i,j,S} \big\{\big|\widehat{\psi}_{ij \mid S}^{[m^*]} - \psi_{ij \mid S}\big|/\sigma(\widehat{\psi}_{ij \mid S})\big\} \leq err_n(M, \nu) \leq \tau(M) \cdot z_{\nu/2L}$ (by assumption 2), when $\psi_{ij \mid S} = 0$, 
\begin{equation*}
    \big|\widehat{\psi}_{ij \mid S}^{[m^*]}\big|/\sigma(\widehat{\psi}_{ij \mid S}) \leq 
    \tau(M) \cdot z_{\nu/2L},
\end{equation*} 
which implies that type I error would not occur if $\mathcal{E}_1$ occurs.

\noindent When $\psi_{ij \mid S} \neq 0$,
\begin{equation*}
    \tau(M) \cdot z_{\nu/2L} \geq err_n(M,\nu) \geq \big|\psi_{ij \mid S} - \widehat{\psi}_{ij \mid S}^{[m^*]}\big|/\sigma(\widehat{\psi}_{ij \mid S}) \geq \big|\psi_{ij \mid S}\big|/\sigma(\widehat{\psi}_{ij \mid S}) - \big|\widehat{\psi}_{ij \mid S}^{[m^*]}\big|/\sigma(\widehat{\psi}_{ij \mid S}),
\end{equation*}  
\begin{equation*}
    \big|\widehat{\psi}_{ij \mid S}^{[m^*]}\big|/\sigma(\widehat{\psi}_{ij \mid S}) \geq \big|\psi_{ij \mid S}\big|/\sigma(\widehat{\psi}_{ij \mid S}) - \tau(M) \cdot z_{\nu/2L}.
\end{equation*}

\noindent Since $\mathop{\lim}\limits_{M \rightarrow \infty}\tau(M) \cdot z_{\nu/L}  = 0$, there exists $M_n$ s.t.
\begin{equation*}
    \tau(M) \cdot z_{\nu/2L} < \big|\psi_{ij \mid S}\big|/4\sigma(\widehat{\psi}_{ij \mid S}) \quad \text{for } M > M_n.
\end{equation*}

\noindent Thus, for $M > M_n$,
\begin{equation*}
    \big|\widehat{\psi}_{ij \mid S}^{[m^*]}\big|/\sigma(\widehat{\psi}_{ij \mid S}) \geq 3\big|\psi_{ij \mid S}\big|/4\sigma(\widehat{\psi}_{ij \mid S}) \geq \tau(M) \cdot z_{\nu/2L},
\end{equation*}
which implies that type II error would not occur if $\mathcal{E}_1$ occurs.\\

\noindent Thus, for $M > M_n$,
\begin{equation*}
    \mathcal{E}_1 \subseteq \Big( \mathop{\cup}\limits_{i,j,S \subseteq \{1,...,d\} \setminus \{i,j\}} \mathcal{E}_{i,j|S} \Big)^c, 
\end{equation*}
where $A^c$ denotes the complement of event $A$.\\

\noindent This implies that for $M > M_n$,
\begin{equation*}
    \mathcal{E}_1 \subseteq \big\{ \widehat{C}^{[m^*]} = C \big\}.
\end{equation*}

\noindent Note that
\begin{equation*}
    \textbf{P}(\beta \in \text{CI}) \geq \textbf{P}(\{ \beta \in \text{CI} \} \cap \mathcal{E}_1) \geq \textbf{P}(\{ \beta \in \text{CI}^{[m^*]} \} \cap \mathcal{E}_1).
\end{equation*}
\begin{equation*}
\begin{split}
    \textbf{P}(\{ \beta \in \text{CI}^{[m^*]} \} \cap \mathcal{E}_1) &= \textbf{P}\Big( \Big\{ \beta \in \mathop{\cup}\limits_{k=1,...,k_{m^*}} \big( \hat{\beta}^{[m^*]}_k - z_{\alpha_1/2}\hat{\sigma}_{\beta_k}^{[m^*]}, \hat{\beta}^{[m^*]}_k + z_{\alpha_1/2}\hat{\sigma}_{\beta_k}^{[m^*]} \big) \Big\} \cap \mathcal{E}_1 \Big) \\
    &\geq \textbf{P} \Big( \Big\{ \beta \in \big(\hat{\beta}^{[m^*]}_{k'} - z_{\alpha_1/2}\hat{\sigma}_{\beta_{k'}}^{[m^*]}, \hat{\beta}^{[m^*]}_{k'} + z_{\alpha_1/2}\hat{\sigma}_{\beta_{k'}}^{[m^*]}\big) \Big\} \cap \mathcal{E}_1 \Big) \\
    &= \textbf{P} \Big( \Big\{ \Big| \frac{\hat{\beta}^{[m^*]}_{k'} - \beta}{\hat{\sigma}_{\beta_{k'}}^{[m^*]}} \Big| \leq z_{\alpha_1/2} \Big\} \cap \mathcal{E}_1 \Big) \quad \text{where } k' \in \{1,...,k_{m^*}\}.
\end{split}
\end{equation*}

\noindent When $\widehat{C}^{[m^*]} = C$, there exists $k' \in \{1,...,k_{m^*}\}$ s.t. $\widehat{G}^{[m^*]}_{k'} = G$ and thus $\hat{\beta}^{[m^*]}_{k'} = \hat{\beta}(\widehat{G}^{[m^*]}_{k'}) = \hat{\beta}(G)$.\\

\noindent Thus, by assumption (3) and Theorem~\ref{theorem1}, we have 
\begin{equation*}
\begin{split}
    \mathop{\liminf}\limits_{n \rightarrow \infty}\mathop{\liminf}\limits_{M \rightarrow \infty} \textbf{P}(\{ \beta \in \text{CI}^{[m^*]} \} \cap \mathcal{E}_1) &\geq \mathop{\liminf}\limits_{n \rightarrow \infty}\mathop{\liminf}\limits_{M \rightarrow \infty} \textbf{P} \Big( \Big\{ \Big| \frac{\hat{\beta}^{[m^*]}_{k'} - \beta}{\hat{\sigma}_{\beta_{k'}}^{[m^*]}} \Big| \leq z_{\alpha_1/2} \Big\} \cap \mathcal{E}_1 \Big) \\
    &= \mathop{\liminf}\limits_{n \rightarrow \infty}\mathop{\liminf}\limits_{M \rightarrow \infty} \textbf{P} \Big( \Big\{ \Big| \frac{\hat{\beta}(G) - \beta}{\hat{\sigma}_\beta} \Big| \leq z_{\alpha_1/2} \Big\} \cap \mathcal{E}_1 \Big) \\
    &= \mathop{\liminf}\limits_{n \rightarrow \infty}\mathop{\liminf}\limits_{M \rightarrow \infty} \Bigg( \textbf{P} \Big( \Big| \frac{\hat{\beta}(G) - \beta}{\hat{\sigma}_\beta} \Big| \leq z_{\alpha_1/2} \Big) - \textbf{P} \Big( \Big\{ \Big| \frac{\hat{\beta}(G) - \beta}{\hat{\sigma}_\beta} \Big| \leq z_{\alpha_1/2} \Big\} \cap \mathcal{E}_1^c \Big) \Bigg) \\
    &\geq \mathop{\liminf}\limits_{n \rightarrow \infty}\mathop{\liminf}\limits_{M \rightarrow \infty} \textbf{P} \Big( \Big| \frac{\hat{\beta}(G) - \beta}{\hat{\sigma}_\beta} \Big| \leq z_{\alpha_1/2} \Big) - \mathop{\liminf}\limits_{n \rightarrow \infty}\mathop{\liminf}\limits_{M \rightarrow \infty} \textbf{P}(\mathcal{E}_1^c) \\
    &\geq (1 - \alpha_1) - \nu \\
    &= 1 - (\gamma - \nu) - \nu \\
    &= 1 - \gamma.
\end{split}
\end{equation*}
\end{proof}

\subsection*{Proof of Theorem~\ref{theorem3}}
\begin{proof}
We first show that, in the limit of large $n$, type II error has a zero probability of occurring in the PC run that produces $\widehat{C}^{[m]}$, for any $1 \leq m \leq M$.
Define the events
\begin{align*}   
    & \mathcal{E}_2 = \Big\{ \big|\widehat{\psi}_{ij \mid S}^{[m]} - \widehat{\psi}_{ij \mid S}\big|/\sigma(\widehat{\psi}_{ij \mid S})  \leq \sqrt{2\log n + 2\log M} \text{ for } 1 \leq m \leq M \Big\},\\
    & \mathcal{E}_3 = \Big\{ \big|\psi_{ij \mid S} - \widehat{\psi}_{ij \mid S}\big|/\sigma(\widehat{\psi}_{ij \mid S})  \leq \sqrt{2\log n} \Big\}.
\end{align*}
Since 
\begin{equation*}
    \mathop{\limsup}\limits_{n \rightarrow \infty} \textbf{P}\big(\big|\widehat{\psi}_{ij \mid S} - \psi_{ij \mid S}\big| / \sigma(\widehat{\psi}_{ij \mid S}) \geq z_{\alpha/2}\big) \leq \alpha \quad \text{for} \quad  0<\alpha<1
\end{equation*}
and
\begin{equation*}
    \widehat{\psi}^{[m]}_{ij \mid S} \overset{\text{i.i.d.}}{\sim} N(\widehat{\psi}_{ij \mid S}, \sigma({\widehat{\psi}_{ij \mid S}})),
\end{equation*}
we obtain
\begin{equation*}
    \mathop{\lim}\limits_{n \rightarrow \infty}\textbf{P}(\mathcal{E}_2 \cap \mathcal{E}_3) = 1
\end{equation*}
by applying the union bound.\\ 

\noindent By the triangular inequality, we have
\begin{equation*}
    \big|\widehat{\psi}^{[m]}_{ij \mid S} - \psi_{ij \mid S}\big|/\sigma({\widehat{\psi}_{ij \mid S}}) \leq \big|\widehat{\psi}^{[m]}_{ij \mid S} - \widehat{\psi}_{ij \mid S}\big|/\sigma({\widehat{\psi}_{ij \mid S}}) + \big|\psi_{ij \mid S} - \widehat{\psi}_{ij \mid S}\big|/\sigma({\widehat{\psi}_{ij \mid S}}).
\end{equation*}
Thus, on $\mathcal{E}_2 \cap \mathcal{E}_3$, we have
\begin{equation}\label{eq 1}
    \big|\widehat{\psi}^{[m]}_{ij \mid S} - \psi_{ij \mid S}\big|/\sigma({\widehat{\psi}_{ij \mid S}}) \leq 2\sqrt{2\log n + 2\log M}.
\end{equation}
Condition \ref{eq:psi low bound} implies that for $\psi_{ij \mid S} \neq 0$,
\begin{equation}\label{eq 2}
    \big| \psi_{ij \mid S} \big|/\sigma({\widehat{\psi}_{ij \mid S}}) > 2\sqrt{2\log n + 2\log M} + 2\log n.
\end{equation}
By combining (\ref{eq 1}) and (\ref{eq 2}), we have, on $\mathcal{E}_2 \cap \mathcal{E}_3$, if $\psi_{ij \mid S} \neq 0$,
\begin{equation*}
    \big|\widehat{\psi}^{[m]}_{ij \mid S} \big|/\sigma({\widehat{\psi}_{ij \mid S}}) > 2\log n \quad \text{for} \quad 1 \leq m \leq M.
\end{equation*}

\noindent Choose a sequence $\nu_n = 2L \times \Big(1 - \Phi \big(2\log(n)/\tau(M)\big)\Big) \to 0$ as $n \to \infty$. Then, the rejection threshold becomes
\begin{equation*}
    \tau(M) \cdot z_{\nu_n/2L} = 2\log n.
\end{equation*}

\noindent Thus, if $\psi_{ij \mid S} \neq 0$,
\begin{equation}\label{eq: type II error}
\begin{split}
    &\mathop{\lim}\limits_{n \rightarrow \infty}\textbf{P}\big(\big|\widehat{\psi}^{[m]}_{ij \mid S} \big|/\sigma({\widehat{\psi}_{ij \mid S}}) > \tau(M)\cdot z_{\nu_n/2L}, \text{ for } 1 \leq m \leq M\big)\\
    &\geq \mathop{\lim}\limits_{n \rightarrow \infty}\textbf{P}(\mathcal{E}_2 \cap \mathcal{E}_3) = 1.
\end{split}
\end{equation}

\noindent Next, we show that in the limit of large $n$, type I error also has a zero probability of occurring in the PC run that produce $\widehat{C}^{[m]}$, for any $1 \leq m \leq M$.\\

\noindent For $1 \leq m \leq M$, 
\begin{equation*}
\begin{split}
    &\textbf{P}\big(\big|\widehat{\psi}_{ij \mid S}^{[m]} - \psi_{ij \mid S}\big|/\sigma(\widehat{\psi}_{ij \mid S}) > \tau(M) \cdot z_{\nu_n/2L}\big)\\ 
    &\leq \textbf{P}\big(\big|\widehat{\psi}_{ij \mid S}^{[m]} - \widehat{\psi}_{ij \mid S}\big|/\sigma(\widehat{\psi}_{ij \mid S}) + \big|\widehat{\psi}_{ij \mid S} - \psi_{ij \mid S}\big|/\sigma(\widehat{\psi}_{ij \mid S}) > \tau(M) \cdot z_{\nu_n/2L}\big)\\
    &\leq \textbf{P}\big(\big|\widehat{\psi}_{ij \mid S}^{[m]} - \widehat{\psi}_{ij \mid S}\big|/\sigma(\widehat{\psi}_{ij \mid S}) > \tau(M)/2 \cdot z_{\nu_n/2L}\big) + \textbf{P}\big(\big|\widehat{\psi}_{ij \mid S} - \psi_{ij \mid S}\big|/\sigma(\widehat{\psi}_{ij \mid S}) > \tau(M)/2 \cdot z_{\nu_n/2L}\big).
\end{split}
\end{equation*}
Since 
\begin{equation*}
    \widehat{\psi}_{ij \mid S}^{[m]} \overset{\text{i.i.d.}}{\sim} N(\widehat{\psi}_{ij \mid S}, \sigma(\widehat{\psi}_{ij \mid S})), \quad 1 \leq m \leq M,
\end{equation*}
we have
\begin{equation*}
    \textbf{P}\big(\big|\widehat{\psi}_{ij \mid S}^{[m]} - \widehat{\psi}_{ij \mid S}\big|/\sigma(\widehat{\psi}_{ij \mid S}) > \tau(M)/2 \cdot z_{\nu_n/2L}\big) = 2 \times \big(1 - \Phi(\tau(M)/2 \cdot z_{\nu_n/2L})\big), 
\end{equation*}
where $\Phi(\cdot)$ is the standard normal c.d.f.\\

\noindent With the choice of $\nu_n = 2L \times \Big(1 - \Phi \big(2\log(n)/\tau(M)\big)\Big) \to 0$ as $n \to \infty$, we have 
\begin{equation}\label{eq: 1st term}
    \mathop{\lim}\limits_{n \rightarrow \infty}\textbf{P}\big(\big|\widehat{\psi}_{ij \mid S}^{[m]} - \widehat{\psi}_{ij \mid S}\big|/\sigma(\widehat{\psi}_{ij \mid S}) > \tau(M)/2 \cdot z_{\nu_n/2L}\big) = 0.
\end{equation}
Since 
\begin{equation*}
    \mathop{\limsup}\limits_{n \rightarrow \infty} \textbf{P}(\big|\widehat{\psi}_{ij \mid S} - \psi_{ij \mid S}\big| / \sigma(\widehat{\psi}_{ij \mid S}) \geq z_{\alpha/2}) \leq \alpha \quad \text{for} \quad  0<\alpha<1,
\end{equation*}
we have
\begin{equation*}
    \mathop{\limsup}\limits_{n \rightarrow \infty} \textbf{P}\big(\big|\widehat{\psi}_{ij \mid S} - \psi_{ij \mid S}\big| / \sigma(\widehat{\psi}_{ij \mid S}) \geq \tau(M)/2 \cdot z_{\nu_n/2L}\big) \leq 2 \times \big(1 - \Phi(\tau(M)/2 \cdot z_{\nu_n/2L})\big).
\end{equation*}
Thus, we obtain
\begin{equation}\label{eq: 2nd term}
    \mathop{\limsup}\limits_{n \rightarrow \infty} \textbf{P}\big(\big|\widehat{\psi}_{ij \mid S} - \psi_{ij \mid S}\big| / \sigma(\widehat{\psi}_{ij \mid S}) \geq \tau(M)/2 \cdot z_{\nu_n/2L}\big) = 0 \quad \text{for any fixed } M.
\end{equation}
Combining (\ref{eq: 1st term}) and (\ref{eq: 2nd term}), if $\psi_{ij \mid S} = 0$, we have
\begin{equation}\label{eq: type I error}
    \mathop{\limsup}\limits_{n \rightarrow \infty}\textbf{P}\big(\big|\widehat{\psi}_{ij \mid S}^{[m]}\big|/\sigma(\widehat{\psi}_{ij \mid S}) > \tau(M) \cdot z_{\nu_n/2L} \text{, for } 1 \leq m \leq M\big) = 0.
\end{equation}

\noindent The combination of (\ref{eq: type II error}) and (\ref{eq: type I error}) implies that 
\begin{equation*}
    \mathop{\limsup}\limits_{n \rightarrow \infty}\textbf{P}\big(\big\{ \widehat{C}^{[m]} = C, 1 \leq m \leq M \big\}\big) = 1
\end{equation*}
and that all $1 \leq m \leq M$ are included in $\mathcal{M}$ as $n \rightarrow \infty$.
Thus,
\begin{equation*}
    \mathop{\limsup}\limits_{n \rightarrow \infty}\textbf{P}(\text{CI}^{\text{re}} = \text{CI}^{\text{oracle}}) = 1.
\end{equation*}
\end{proof}

\section*{Appendix 2: Additional simulation results}

\subsection*{Appendix 2.1: Sparse DAG}\label{app2.1}
\begin{figure}
\centerline{\includegraphics[width=40 pc,height=32 pc]
{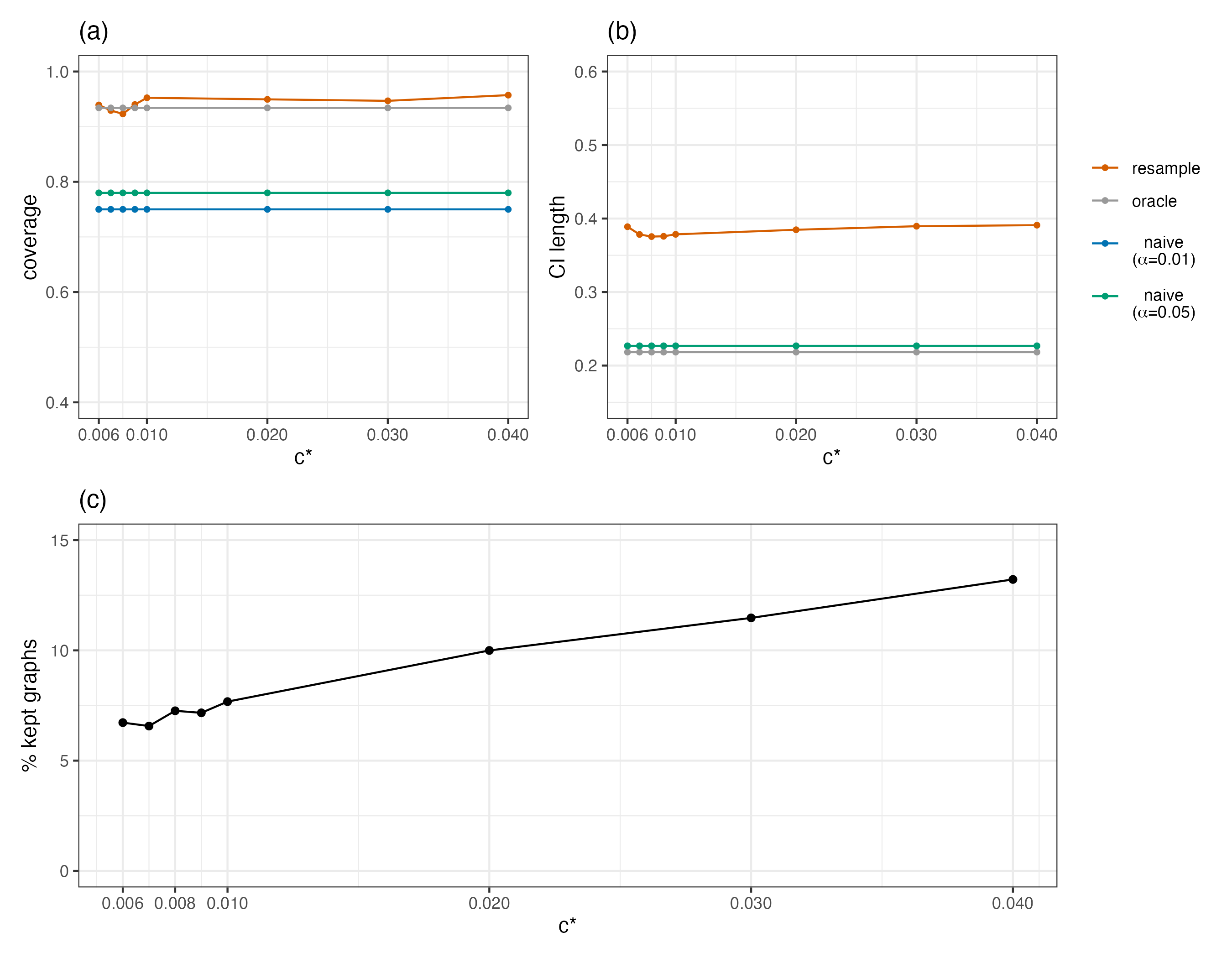}}
\caption{(a) Empirical coverage rate and (b) average length of $95\%$ CIs based on 500 simulations for varying $c^*$ values. ``naive" is the CI without incorporating the uncertainty in causal graph selection, ``oracle" is the CI with knowledge of the true causal graph, and ``resample" is our proposed CI in Section~\ref{sec3}, with $M=50$ resamples. (c) Percentage of valid graphs with $M=100$ resamples for constructing our proposed CI; results are based on 500 simulations for varying $c^*$. The true graph model possesses an average of 4 neighbors per node.}
\label{supp_sparse}
\end{figure}

We perform additional simulations following the data generation procedure detailed in Section 5.1, except that the expected number of neighbors per node is now set to 4 to generate a sparse graph. We let the sample size $n = 500$, resampling size $M = 50$, $\text{max}_{i \in V} |\text{Adj}_i(G)| = 7$ and $\nu = 0.025$. As in Section 5.2, we consider $c^* \in \{0.006, 0.007, 0.008, 0.009, 0.01, 0.02, 0.03, 0.04\}$ and run 500 simulations for each $c^*$. Figure~\ref{supp_sparse} shows the empirical coverage rate and $95\%$ CI length from different methods, along with the percentage of valid graphs from the resampling method. Most notably, in comparison to Figure~\ref{fig1}, we find that the coverage rate of $\text{CI}^{\text{re}}$ in this case is more robust to the choice of $c^*$. $\text{CI}^{\text{re}}$ maintains nominal coverage even with $c^*$ as large as 0.04, suggesting that our method generally performs well in the case of learning sparse DAGs. The CI length and the percentage of kept graphs follow similar patterns as in the case with dense graphs.

\subsection*{Appendix 2.2: Truncated resampling}\label{app2.2}
\begin{figure}
\centerline{\includegraphics[width=40 pc,height=32 pc]
{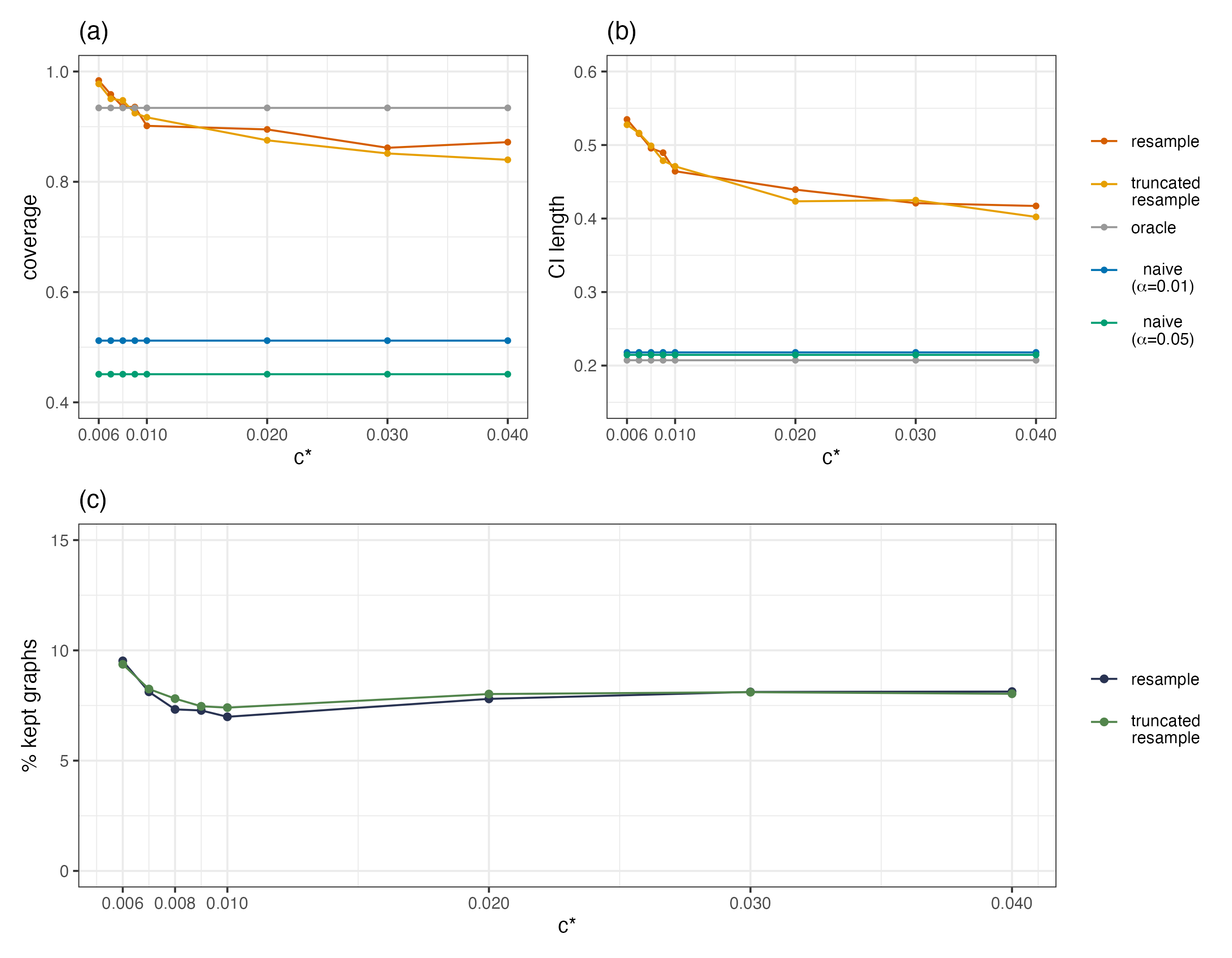}}
\caption{(a) Empirical coverage rate and (b) average length of $95\%$ CIs based on 500 simulations for varying $c^*$ values. ``naive" is the CI without incorporating the uncertainty in causal graph selection, ``oracle" is the CI with knowledge of the true causal graph, ``resample" is our proposed CI in Section~\ref{sec3}, with $M=50$ resamples, and ``truncated resample" is the same as ``resample", but with resampling of test statistics from a truncated Gaussian distribution. (c) Percentage of valid graphs with $M=100$ resamples for constructing our proposed CI; results are based on 500 simulations for varying $c^*$. The true graph model possesses an average of 7 neighbors per node.}
\label{supp_trunc}
\end{figure}

Resampling test statistics from a Gaussian distribution can theoretically produce resamples that are distant from the mean, i.e., the sample test statistic. To assess the empirical sensitivity of $\text{CI}^{\text{re}}$ to resamples from the tails of the Gaussian distribution, we include in Figure~\ref{supp_trunc}, on top of the results in Figure~\ref{fig1}, a version of our proposed approach where the resampled test statistics $Z(\hat{\rho}^{[m]}_{ij \mid S})$ are drawn from a Gaussian distribution truncated at $\pm 1.5$ standard deviations from the sample test statistic $Z(\hat{\rho}_{ij \mid S})$. Given the similar results from both resampling-based CIs, Figure~\ref{supp_trunc} suggests that in practice, the performance of $\text{CI}^{\text{re}}$ should not be significantly influenced by potential resamples from the tails.

\subsection*{Appendix 2.3: Bootstrap procedure}\label{app2.3}
\begin{figure}
\centerline{\includegraphics[width=30 pc,height=20pc]
{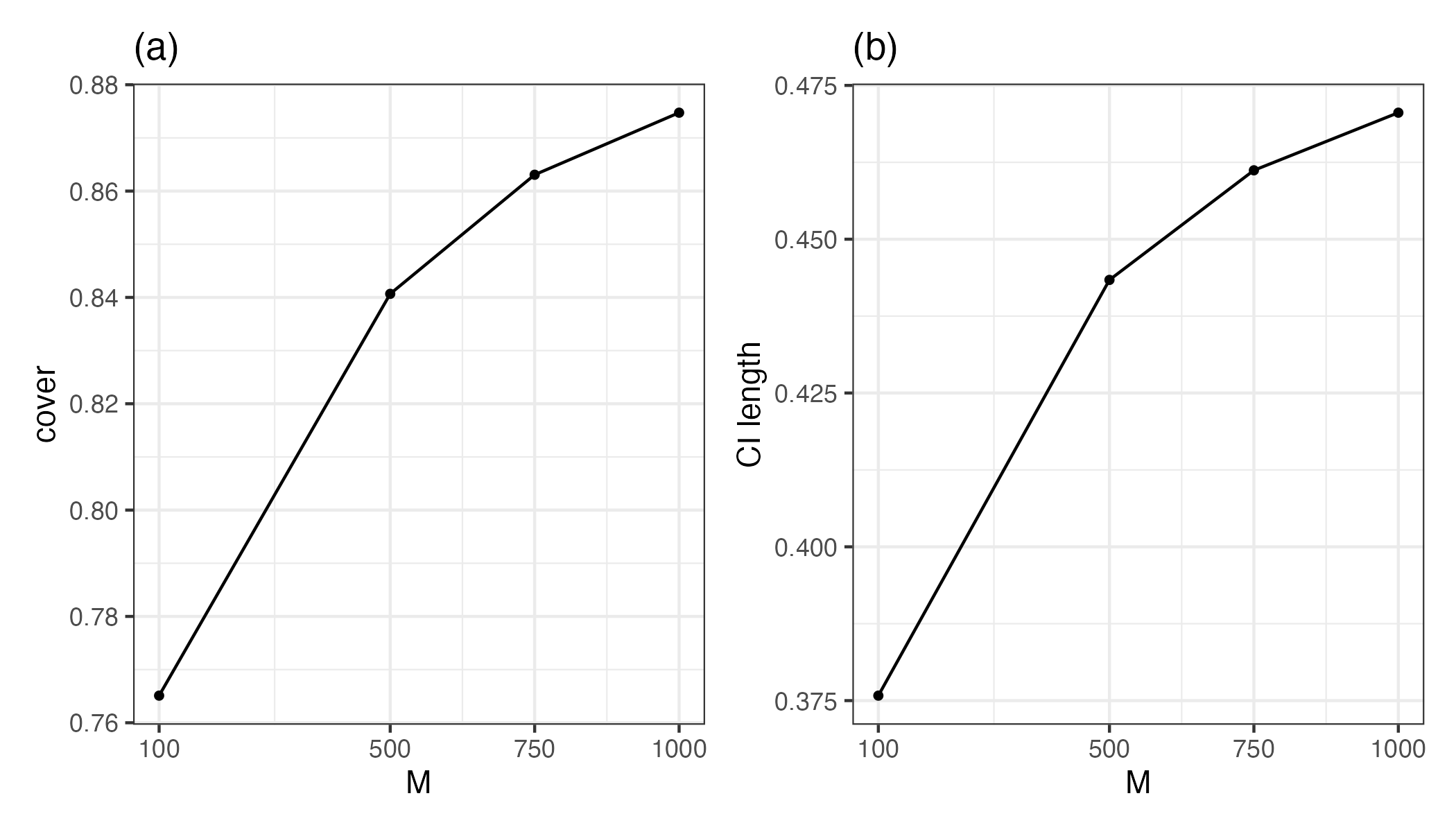}}
\caption{(a) Empirical coverage rate and (b) average length of 95\% CIs based on 500 simulations for 100, 500, 750, and 1,000 repetitions of bootstrap resampling. The true graph model possesses an average of 7 neighbors per node.}
\label{supp_boot}
\end{figure}

The PC-algorithm relies on a series of hypothesis tests, and it has been shown that using test statistics from bootstrap samples can lead to inflated type I errors and invalid p-values \citep{janitza2016pitfalls}. As a result, a set of graphs esimated from bootstrap samples will not necessarily include the true graph. We perform simulations to evaluate the performance of a nonparametric bootstrap procedure (Algorithm~\ref{alg boot}) under the setting described in Section 5.1. Similar to our proposed approach, we construct a CI (estimated on the \emph{original} sample) for each of the estimated graphs that is a valid CPDAG, and then take the union of these CIs. We let the sample size $n=500$ and $M = 100, 500, 750, 1000$. Figure~\ref{supp_boot} shows the empirical coverage rate and 95\% CI length after 500 simulation runs. Compared to our proposed procedure, the nonparametric bootstrap procedure yield poorer coverage rates and similar CI lengths. With 1000 bootstrap samples, the coverage rate is 87\%, whereas our proposed approach achieves a coverage rate of 90\% with only $M=50$ ($c^*=0.01$).\\

\begin{algorithm}[H]
\caption{Executing the PC-algorithm multiple times using bootstrap samples. }\label{alg boot}
\begin{algorithmic}[1]
\Require Samples of the vector $X = (X_1,...,X_d)'$ and $\text{O}(V)$
\Ensure $M$ graphs 
\For {$m = 1,...,M$}
    \State Resample data with replacement from the original sample
    \State Do Algorithm~\ref{alg1} ($\alpha=0.01$) on the resampled data
    \State \Return $\widehat{C}^{[m]}$
\EndFor
\State \Return $(\widehat{C}^{[1]},...,\widehat{C}^{[M]})$
\end{algorithmic}
\end{algorithm}

\subsection*{Appendix 2.4: Nonparametric independence test}\label{app2.4}
The PC-algorithm can be applied using nonparametric tests of conditional independence, such as one based on the generalized covariance measure (GCM) by \citet{shah2020hardness}. To introduce the GCM test statistic, consider $n$ i.i.d. observations of a random triple $(X,Y,Z)$, where $(X,Y,Z)$ follows a joint distribution $P$ that is absolutely continuous with respect to Lebesgue measure. The objective is to test whether $X$ and $Y$ are conditionally independent given $Z$. Let $f_P(z) = E_P[X|Z=z]$ and $g_P(z) = E_P[Y|Z=z]$, with corresponding nonparametric regression estimates $\hat{f}$ and $\hat{g}$. For $k=1,...,n$, let $R_k$ be the product between residuals:
\begin{equation*}
    R_k = \{x_k-\hat{f}(z_k)\}\{y_k-\hat{g}(z_k)\}.
\end{equation*}
The test statistic is defined as \citep{shah2020hardness}
\begin{equation*}
    T = \frac{\sqrt{n} \cdot \frac{1}{n}\sum_{k=1}^nR_k}{\Big(\frac{1}{n}\sum_{k=1}^n R_k^2 - \big(\frac{1}{n}\sum_{k=1}^n R_k\big)^2 \Big)^{1/2}} =: \frac{\tau_N}{\tau_D}.
\end{equation*}
Large values of $|T|$ would reject the null of conditional independence. Define 
\begin{equation*}
    \rho_P = E_P[\text{cov}_P(X,Y|Z)].
\end{equation*}
Under the conditions in Theorems 6 and 8 of \citet{shah2020hardness}, 
\begin{equation*}
    \frac{|\tau_N - \sqrt{n}\rho_P|}{\tau_D} \to_d N(0,1).
\end{equation*}
Thus, GCM meets the asymptotic normality condition in Theorem~\ref{theorem1}. We can draw $\hat{\rho}$ from $N(\tau_N/\sqrt{n}, \tau_D/\sqrt{n})$ to generate resampled test statistics, specifically, the $m$-th resampled test statistic is 
\begin{equation*}
    T^{[m]} = \frac{\sqrt{n}\hat{\rho}^{[m]}}{\tau_D}.
\end{equation*}

We perform simulations to evaluate the performance of our proposed approach with conditional independence tests based on the GCM. Due to the high computational cost of nonparametric tests, we generate a small graph -- a 5-node random DAG $G$ where the expected number of neighbors per node is 3 -- with the remainder of the data generating process following that described in Section 5.1. We assume knowledge of the temporal order O$(V)=(1,2,2,2,3)$; the target estimand is the average causal effect of variable 4 on variable 5, $\beta_{4,5}(G)$. We set the sample size to $n=500$, the maximum node degree to 3, and choose $\nu=0.025$, $c^*=0.01$, and $M=20$. Residuals for the GCM test are computed using the \texttt{R} package \texttt{GeneralisedCovarianceMeasure}, with regressions $\hat{f}$ and $\hat{g}$ estimated via XGBoost \citep{gcm}. Based on 500 simulation runs, the empirical coverage rate of $\text{CI}^{\text{re}}$ is 100\% (compared to 71.9\% for $\text{CI}^{\text{naive}}$ with $\alpha=0.01$), with an average 95\% CI length of 0.41 (versus 0.24 for the naive approach).

\section*{Appendix 3: Remarks on the selection of the resampling size}
Following the proof of Theorem~\ref{theorem2}, we assume the event 
    \begin{equation*}
        \mathcal{E}_1 = \Big\{ \mathop{\min}\limits_{1 \leq m \leq M}\mathop{\max}\limits_{i,j,S} \big\{ \big|\widehat{\psi}_{ij \mid S}^{[m]} - \psi_{ij \mid S}\big|/\sigma(\widehat{\psi}_{ij \mid S}) \big\} \leq err_n(M, \nu) \Big\}
        \end{equation*}
holds (as well as assumptions of Theorem~\ref{theorem2}) and we let $m^*$ denote the index s.t.
    \begin{equation*}
        \mathop{\max}\limits_{i,j,S} \big\{\big|\widehat{\psi}_{ij \mid S}^{[m^*]} - \psi_{ij \mid S}\big|/\sigma(\widehat{\psi}_{ij \mid S})\big\} \leq err_n(M, \nu).
    \end{equation*}
Then, for any fixed $n$, there exists $M_n$ s.t. for $M>M_n$, the true CPDAG is recovered (i.e., $\widehat{C}^{[m^*]}=C$). The expression for $M_n$ can be obtained by solving (for some $\psi_{ij \mid S} \neq 0$)
    \begin{equation*}
        \tau(M) \cdot z_{\nu/2L} < \big|\psi_{ij \mid S}\big|/4\sigma(\widehat{\psi}_{ij \mid S})
    \end{equation*}
where $\tau(M)=c^* \cdot (\log n/M)^{1/L}$. By substituting $\sigma(\widehat{\psi}_{ij \mid S})$ with $\mathop{\max}\limits_{i,j,S}\{\sigma(\widehat{\psi}_{ij \mid S})\}$ and $|\psi_{ij \mid S}|$ with $\mathop{\min}\limits_{i,j,S}|\{ \psi_{ij \mid S} : \psi_{ij \mid S}>0 \}|$, we have the following expression
    \begin{equation*}
        M_n = \Bigg[\frac{4c^* z_{\nu/2L}\cdot \mathop{\max}\limits_{i,j,S}\{\sigma(\widehat{\psi}_{ij \mid S})\}}{\mathop{\min}\limits_{i,j,S}|\{ \psi_{ij \mid S} : \psi_{ij \mid S}>0 \}|}\Bigg]^L \cdot \log n.
    \end{equation*}
Since $M_n$ depends on several unknown quantities, it is a purely theoretical and conservative approach to choosing $M$. The argument here closely resembles what \citet{guo2023robust} have in their Lemma 1.

So, the theoretical statement we can make is the following: if we assume $M$ and $n$ are large enough such that $\mathcal{E}_1$ holds, then choosing $M$ according to this formula is sufficient such that the true CPDAG is recovered in one of the resamples (and hence the coverage guarantee is met). However, we do not know in general that $M$ and $n$ are large enough such that $\mathcal{E}_1$ holds, we only know that this event holds with high probability as $M,n \to \infty$. But interestingly this assumption is just about our ability to estimate the true conditional associations with low error, which is statistically ``easier'' than model selection.

\newpage
\bibliographystyle{unsrtnat}
\bibliography{references}
\end{document}